\documentclass[11pt]{article}

\title{Anisotropic Brown-Resnick space-time processes: \\estimation and model assessment}

\author{Sven Buhl\thanks{Center for Mathematical Sciences, Technische Universit\"at M\"unchen,  85748 Garching, Boltzmannstrasse 3, Germany, email: sven.buhl@tum.de\,,\,cklu@tum.de, url: {http://www.statistics.ma.tum.de}}
\and
Claudia Kl\"uppelberg\footnotemark[1]
}

\usepackage{amssymb}
\usepackage{amsfonts}
\usepackage{amsmath}
\usepackage{amsthm}
\usepackage[numbers]{natbib}
\usepackage{comment}
\usepackage{color}
\usepackage[font={footnotesize}]{caption}
\usepackage[markup=ncolor]{changes}
\colorlet{Changes@Color}{red}
\usepackage{ulem}
\usepackage{bbm}
\usepackage{latexsym}
\usepackage{float}
\usepackage{amssymb}
\usepackage{amsmath}
\usepackage{mathrsfs}
\usepackage{graphicx}

\usepackage{paralist}
\usepackage{url}
\usepackage{float}
\usepackage[caption=false]{subfig}
\numberwithin{equation}{section}

\newtheorem{theorem}{Theorem}
\newtheorem{lemma}{Lemma}
\newtheorem{remark}{Remark}
\newtheorem{example}{Example}
\newtheorem{proposition}{Proposition}
\newtheorem{definition}{Definition}
\newtheorem{corollary}{Corollary}

\newtheorem{fig}{Figure}

\newcommand{\bthe}{\begin{theorem}}
\newcommand{\ethe}{\end{theorem}}

\newcommand{\ben}{\begin{enumerate}}
\newcommand{\een}{\end{enumerate}}

\newcommand{\bit}{\begin{itemize}}
\newcommand{\eit}{\end{itemize}}

\newcommand{\ex}{\textnormal{e}^}

\newcommand{\beq}{\begin{equation}}
\newcommand{\eeq}{\end{equation}}

\newcommand{\ble}{\begin{lemma}}
\newcommand{\ele}{\end{lemma}}

\newcommand{\bde}{\begin{definition}\rm}
\newcommand{\ede}{\halmos\end{definition}}

\newcommand{\bco}{\begin{corollary}}
\newcommand{\eco}{\end{corollary}}

\newcommand{\bpr}{\begin{proposition}}
\newcommand{\epr}{\end{proposition}}

\newcommand{\brem}{\begin{remark}\rm}
\newcommand{\erem}{\end{remark}}

\newcommand{\bproof}{\begin{proof}}
\newcommand{\eproof}{\end{proof}}

\newcommand{\bexam}{\begin{example}\rm}
\newcommand{\eexam}{\end{example}}

\newcommand{\bfi}{\begin{fig}}
\newcommand{\efi}{\end{fig}}

\newcommand{\btab}{\begin{tab}}
\newcommand{\etab}{\end{tab}}

\newcommand{\beao}{\begin{eqnarray*}}
\newcommand{\eeao}{\end{eqnarray*}\noindent}

\newcommand{\beam}{\begin{eqnarray}}
\newcommand{\eeam}{\end{eqnarray}\noindent}

\newcommand{\barr}{\begin{array}}
\newcommand{\earr}{\end{array}}

\newcommand{\bdis}{\begin{displaymath}}
\newcommand{\edis}{\end{displaymath}\noindent}

\newcommand{\R}{\mathbb R}
\newcommand{\N}{\mathbb N}

\renewcommand{\P}{\mathbb P}
\newcommand{\E}{\mathbb E}

\newcommand{\cals}{\mathcal{S}}

\newcommand{\calf}{\mathcal{F}}

\newcommand{\dsum}{\displaystyle\sum}

\newcommand{\limd}{\stackrel{\mathcal{D}}{\rightarrow}}
\newcommand{\stp}{\stackrel{P}{\rightarrow}}
\newcommand{\std}{\stackrel{\mathcal{D}}{\rightarrow}}
\newcommand{\stas}{\stackrel{\rm a.s.}{\rightarrow}}

\newcommand{\eqd}{\stackrel{d}{=}}

\newcommand{\nto}{n\to\infty}

\newcommand{\al}{{\alpha}}

\newcommand{\var}{{\mathbb{V}ar}}
\newcommand{\cov}{{\mathbb{C}ov}}

\newcommand{\wh}{\widehat}

\newcommand{\bs}{\boldsymbol}

\newcommand{\halmos}{\quad\hfill\mbox{$\Box$}}  %end of proof sign

\allowdisplaybreaks[4]

\begin{document}

%\date{}

\maketitle

\begin{abstract}
Spatially isotropic max-stable processes have been used to model extreme spatial or space-time observations. One prominent model is the Brown-Resnick process, which has been successfully fitted to time series, spatial data and space-time data. This paper extends the process to possibly anisotropic spatial structures. For regular grid observations we prove strong consistency and asymptotic normality of pairwise maximum likelihood estimates for fixed and increasing spatial domain, when the number of observations in time tends to infinity. We also present a statistical test for isotropy versus anisotropy. We apply our test to precipitation data  in Florida, and present some diagnostic tools for model assessment.  Finally, we present a method to predict conditional probability fields and apply it to the data.
\end{abstract}



\section{Introduction}\label{s1}

Max-stable processes, such as the Brown-Resnick process, have been successfully fitted to time series, spatial and recently to space-time data. Methods for inference include pairwise likelihood based on the bivariate density of the models (cf. \citet{Ribatet}), censored likelihood (cf. \citet{Wadsworth2}) or threshold-based approaches (cf. \citet{Engelke1}).
In \citet{Steinkohl2} a spatially isotropic Brown-Resnick space-time process is suggested and applied to  precipitation data.
Pairwise maximum likelihood estimates are shown to be strongly consistent and asymptotically normal, provided the domain of observations increases jointly in space and time.
Their approach is restricted to isotropic spatial dependence.

In the present paper we generalise the Brown-Resnick model to allow anisotropy in space. 
The new model allows for different extremal behaviour along orthogonal spatial directions. 
Anisotropy is often observed on Earth, for example in Middle Europe with its westerly winds or near the equator where trade winds involve predominant easterlies.
All dependence parameters are summarised in the semivariogram of an underlying Gaussian space-time process.
This semivariogram then defines the dependence structure of the max-stable process and, as a consequence, the tail dependence coefficient between two process values evaluated at two location and two time points. 

Furthermore, since in real world applications, observations are often recorded over a large number of time points, but only at a comparably small number of spatial locations, we  consider both a fixed and increasing spatial domain in combination with an increasing temporal domain.
For both settings, fixed and increasing spatial domain, we prove strong consistency and asymptotic normality of the pairwise maximum likelihood estimates in the anisotropic model based on regular grid observations.
This requires in particular to prove space-time and temporal mixing conditions in both settings for the anisotropic model.

We also provide tests for isotropy versus anisotropy again in both settings, which are designed for the new model.
The asymptotic normality of the parameter estimates determines in principle the rejection areas of the test. 
However, the covariance matrices of the normal limit laws are not available in closed form. 
We formulate a subsampling procedure in the terminology of the Brown-Resnick space-time process and prove its convergence for fixed and increasing spatial domain. 

We conclude with an analysis of space-time block maxima of radar rainfall measurements in Florida. Firstly, we present a simple procedure to test whether they originate from a max-stable process.
As this cannot be rejected, we fit the Brown-Resnick space-time model to the data, using pairwise maximum likelihood estimation. Subsequently we apply the new isotropy test. Both the estimation and the test are based on the setting of a fixed spatial domain and increasing time series.
In particular, since the Brown-Resnick space-time process satisfies the strong mixing conditions for increasing spatial and time domain as well as for fixed spatial and increasing time domain, the estimation and test procedure are independent of the specific setting: it works in both settings in exactly the same way, taking the different asymptotic covariance matrices into account.
Finally, we assess the goodness of fit of the estimated model by a simulation diagnostics based on a large number of i.i.d. simulated anisotropic Brown-Resnick space-time processes. 
As a result, there is no statistical significance that the anisotropic Brown-Resnick space-time process with the fitted parameters should be rejected.

Our paper is organised as follows. In Section~\ref{s2} we present the Brown-Resnick space-time model, which allows for anisotropic effects in space, and various dependence measures, including the parameterised dependence function. 
In Section~\ref{s3} we compute the pairwise maximum likelihood estimates for the new model and prove their strong consistency and asymptotic normality for both settings, fixed and increasing spatial domain. 
Section~\ref{s4} presents hypothesis tests for spatial isotropy and derives rejection areas based on a subsampling procedure.
A data analysis is performed in Section~\ref{s5} with focus on model assessment.
The isotropy test rejects spatial isotropy for these data in favour of our new anisotropic model.
Based on two other test procedures, we conclude that the anisotropic Brown-Resnick space-time process with the given dependence parameters is an appropriate model for the block-maxima data. 
We conclude by predicting conditional probability fields, which give the probability of a high value (for example of the amount of precipitation) at some space-time location given a high  value at some other location.

\section{Spatially anisotropic Brown-Resnick processes}\label{s2}

Throughout the paper we consider a \textit{stationary Brown-Resnick space-time process} with representation
\begin{align}\label{BR}
\eta(\bs{s},t) = \bigvee\limits_{j=1}^\infty \left\{\xi_j \,  e^{W_j(\bs{s},t)-\delta(\bs{s},t)} \right\},\quad (\bs{s},t)\in\R^d\times [0,\infty),
\end{align}
where 
$\{\xi_j : j\in\N\}$ are points of a Poisson process on $[0,\infty)$ with intensity $\xi^{-2}d\xi$, the \textit{ dependence function} $\delta$ is nonnegative and conditionally negative definite and 
$\{W_j(\bs{s},t): \bs{s} \in \mathbb{R}^d, t \in [0, \infty)\}$ are independent replicates of a Gaussian process \\ $\{W(\bs s,t): \bs{s} \in \mathbb{R}^d, t \in [0, \infty)\}$ with stationary increments,
$W(\bs{0},0)=0$, $\E [W(\bs{s},t)]=0$ and covariance function
$$\cov[W(\bs{s}^{(1)},t^{(1)}),W(\bs{s}^{(2)},t^{(2)})] 
=\delta(\bs{s}^{(1)},t^{(1)})+\delta(\bs{s}^{(2)},t^{(2)}) -\delta(\bs{s}^{(1)}-\bs{s}^{(2)}, t^{(1)}-t^{(2)}).$$
Representation \eqref{BR} goes back to \citet{deHaan} and \citet{Gine}. 
 Brown-Resnick processes have been studied by \citet{Brown} in a time series context, as a spatial model by \citet{Schlather2}, and in a space-time setting by  \citet{Steinkohl} and \citet{Huser}.
 The univariate margins of the process $\eta$ follow standard Fr{\'e}chet distributions. 

There are various quantities to describe the dependence in \eqref{BR}:
\begin{itemize}
\item 
In geostatistics, the {dependence function} $\delta$ is termed the \textit{semivariogram} of the process $\{W(\bs s, t)\}$: For $(\bs{s}^{(1)},t^{(1)}), (\bs{s}^{(2)},t^{(2)}) \in \R^d\times[0,\infty)$, it holds that $$\var[W(\bs{s}^{(1)},t^{(1)})-W(\bs{s}^{(2)},t^{(2)})]=2\delta(\bs{s}^{(1)}-\bs{s}^{(2)},t^{(1)}-t^{(2)}).$$
\item 
For $\bs{h} \in \mathbb{R}^d$ and $u \in \mathbb{R}$, the \textit{tail dependence coefficient} $\chi(\bs{h},u)$ is given by (cf. \citet{Schlather2}, Remark~25 or \citet{Steinkohl}, Section~3)
\begin{align}\label{taildependencecoeffbrownresnick}
\hspace*{-6mm}\chi(\bs{h},u):=\lim\limits_{y \rightarrow \infty} \mathbb{P} \big(\eta(\bs{s}^{(1)},t^{(1)})>y \mid \eta(\bs{s}^{(2)},t^{(2)})>y\big)
= 2\bigg(1-\Phi\bigg({\sqrt{\frac{\delta(\bs{h},u)}{2}}}\bigg)\bigg),
\end{align}\noindent
where $\bs{h}=\bs{s}^{(1)}-\bs{s}^{(2)}$, $u=t^{(1)}-t^{(2)}$, and $\Phi$ denotes the standard normal distribution function.
\item 
For $D=\{(\bs{s}^{(1)},t^{(1)}),\ldots,(\bs{s}^{(|D|)},t^{(|D|)})\} $ and $\bs{y}=(y_1,\ldots,y_{|D|})>\bs 0$ the finite-dimen\-sional margins are given by
\begin{align}
\P(\eta(\bs{s}^{(1)},t^{(1)})\le y_1, \eta(\bs{s}^{(2)},t^{(2)})\le y_2,\ldots, \eta(\bs{s}^{(|D|)},t^{(|D|)})\le y_{|D|}) =  e^{-V_D(\bs y)}. \label{exponentmeasureD}
\end{align}
Here $V_D$ denotes the \textit{exponent measure}, which is homogeneous of order -1. 
\item 
The \textit{extremal coefficient} $\xi_D$ for any finite set $D\subset\R^{d}\times [0,\infty)$ is defined through
$$\P(\eta(\bs{s}^{(1)},t^{(1)})\le y, \eta(\bs{s}^{(2)},t^{(2)})\le y,\ldots, \eta(\bs{s}^{(|D|)},t^{(|D|)})\le y) =  e^{-\xi_D/y},\quad y>0;$$
i.e., $\xi_D=V_D(1,\ldots,1).$ If $|D|=2$, then (cf. \citet{Beirlant}, Section 9.5.1) 
$$\chi(\bs{s}^{(1)}-\bs{s}^{(2)},t^{(1)}-t^{(2)}) = 2-\xi_D.$$ 
\end{itemize}

In this paper we assume the dependence function $\delta$ to be given for spatial lag $\bs h$ and time lag $u$ by
\begin{align}\label{vario0}
\delta(\bs h,u) = \sum_{j=1}^d C_j |h_j|^{\alpha_j}+C_{d+1} |u|^{\alpha_{d+1}}, \quad (\bs h,u)=(h_1,\ldots, h_d,u) \in \mathbb{R}^{d+1},
\end{align}\noindent 
with parameters $C_j>0$ and $\alpha_j \in (0,2]$ for $j=1,\ldots,d+1$. \\
Model~\eqref{vario0} allows for different rates of decay of extreme dependence in different directions. This particularly holds along the axes of a $d$-dimensional spatial grid, but also for other directions. For example in the case $d=2$, the decreases of dependence along the directions $(1,2)$ and $(2,1)$ differ. Model~\eqref{vario0} can be generalised by a simple rotation to a setting, where not necessarily the axes, but other principal orthogonal directions play the major role. The rotation angle then needs to be estimated together with the other model parameters. 
A similar approach has been applied to introduce geometric or zonal anisotropy into a spatial isotropic model (see e.g. \citet{Blanchet1}, Section~4.2, or \citet{Engelke1}, Section~5.2). 
For a justification of model \eqref{vario0} see \citet{Buhlma}, Sections~4.1 and~4.2. 
There it is shown that Brown-Resnick processes with this dependence function arise as limits of appropriately rescaled maxima of Gaussian processes with a large variety of correlation functions. 

\section{Pairwise maximum likelihood estimation}\label{s3}

We extend the pairwise maximum likelihood procedure described in \citet{Steinkohl2} for spatially isotropic space-time Brown-Resnick processes to the anisotropic case. 
We focus on the difference introduced by the spatial anisotropy and refer to the corresponding formulas in \citet{Steinkohl2}, where also a short introduction to composite likelihood estimation and further references can be found.

The pairwise likelihood function uses the bivariate distribution function of \linebreak
$(\eta(\bs{s},t), \eta(\bs{s}+\bs h,t+u))\eqd (\eta(\bs{0},0), \eta(\bs{h},u))$ (equal in distribution by stationarity) for $\bs{h}\in\R^d$ and $u\in\R$, which is given as
\begin{align}\label{pairlike}
G(y_1,y_2) = \exp\{-V(y_1,y_2)\},\quad y_1,y_2>0,
\end{align}\noindent
where the exponent measure $V=V_{D}$ for {$D=\{(\bs{s}^{(1)},t^{(1)}),(\bs{s}^{(2)},t^{(2)})\}$}  has the representation
\begin{align}
\lefteqn{V(y_1,y_2)}\nonumber \\
&=& \frac{1}{y_1}\Phi\left(\frac{\log(y_2/y_1)}{{\sqrt{2\delta(\bs{h},u)}}}+{\sqrt{\frac{\delta(\bs{h},u)}{2}}}\right)
+\frac{1}{y_2}\Phi\left(\frac{\log(y_1/y_2)}{{\sqrt{2\delta(\bs{h},u)}}}+{\sqrt{\frac{\delta(\bs{h},u)}{2}}}\right), \label{exponentmeasure}
\end{align}\noindent
which is a particular form of Eq.~(2.7) in \citet{Huesler}.
The dependence function $\delta$ is given by  \eqref{vario0}.
For a derivation of \eqref{exponentmeasure} see for instance \citet{Oestingdipl}, Satz und Definition~2.4. 

From this we can calculate the pairwise density $g(y_1,y_2)=g_{\bs{\theta}}(y_1,y_2)$ of $G$ by differentiation. The parameter  vector $\bs{\theta}=(C_1,\ldots,C_{d+1},\alpha_1,\ldots,\alpha_{d+1})$ lies in the parameter space 
$$\Theta:=\left\{(C_1,\ldots,C_{d+1},\alpha_1,\ldots,\alpha_{d+1}): C_j \in (0, \infty), \alpha_j \in (0,2], j=1,\ldots,d+1\right\}.$$

We focus on data on a regular spatial grid and at equidistant time points. 
More precisely, we assume that the spatial observations lie on a regular $d$-dimensional lattice,
$$\cals_M = \{\bs s=(s_1,\ldots,s_d)\in\{1,\ldots,M\}^d\}$$
for $M\in\N$, and that the time points are given by the set $\mathcal{T}_T=\{1,\ldots,T\}$ for $T\in\N$.

For the computation of the pairwise likelihood it is common not to include observations on all available space-time pairs, but only on those that lie within some prespecified spatio-temporal distance. This is motivated by the fact that pairs which lie sufficiently far apart in a space-time sense have little influence on the dependence parameters, see \citet{Nott}, Section~2.1. To express this notationally, we take inspiration from that paper and use a design mask adapted to the anisotropic setting, 
\begin{align}
\mathcal{H}_{\bs{r}}:=\big\{\bs{h}=(h_1,\ldots, h_d) \in \mathbb{N}_0^d: \bs{h} \leq \bs{r}\big\}, \quad \bs{r}=(r_1, \ldots, r_d) \in \mathbb{N}_0^d. \label{designmaskaniso}
\end{align}
We are now ready to define the pairwise log-likelihood function and the resulting estimate.

\begin{definition}[Pairwise likelihood estimate]
The pairwise log-likelihood function based on space-time pairs, whose maximum spatial lag is $\bs{r} \in \mathbb{N}_0^d$ and maximum time lag is $p \in \mathbb{N}_0$, such that $(\bs r, p) \neq (\bs 0,0)$, is defined as 
\begin{align}
PL^{(M,T)}(\bs{\theta})&:={\sum\limits_{\bs s \in \mathcal{S}_M}} \sum\limits_{t=1}^T \sum\limits_{\bs{h} \in \mathcal{H}_{\bs{r}} \atop \bs s+ \bs{h}\in\cals_M} \sum\limits_{u=0 \atop t+u \leq T}^p \mathbbmss{1}_{\{(\bs h, u) \neq (\bs 0,0)\}}\log \left\{g_{\bs{\theta}}\left(\eta(\bs s,t), \eta(\bs s+\bs{h}, t+u) \right) \right\} \notag\\
&={\sum\limits_{\bs s \in \mathcal{S}_M}} \sum\limits_{t=1}^T q_{\bs{\theta}}(\bs s,t; \bs{r},p) -\mathcal{R}^{(M,T)}(\bs{\theta}), \quad \bs{\theta} \in \Theta, \label{PWLgeneraldimension}
\end{align}
where
\begin{align}
q_{\bs{\theta}}(\bs s, t; \bs{r},p):=\sum\limits_{\bs{h} \in \mathcal{H}_{\bs{r}}} \sum\limits_{u=0}^p \mathbbmss{1}_{\{(\bs h, u) \neq (\bs 0,0)\}} \log \left\{g_{\bs{\theta}}\left(\eta(\bs s,t), \eta(\bs s+\bs{h}, t+u)\right)\right\} \label{qteta}
\end{align}
and 
\begin{align} 
\mathcal{R}^{(M,T)}(\bs{\theta})  
&:={\sum\limits_{\bs s \in \mathcal{S}_M}} \sum\limits_{t=1}^T \sum\limits_{\bs{h} \in \mathcal{H}_{\bs{r}}} \sum\limits_{u=0}^p \mathbbmss{1}_{\{\bs s+ \bs{h}\notin\cals_M \textnormal{ or } t+u > T \}} \log \left\{g_{\bs{\theta}}\left(\eta(\bs s,t), \eta(\bs s+\bs{h}, t+u) \right) \right\} \nonumber\\
&=\sum\limits_{\bs{h} \in \mathcal{H}_{\bs{r}}} \sum\limits_{u=0}^p \sum\limits_{(\bs s,t) \in \mathcal{G}_{M,T}(\bs{h},u)} \log\left\{g_{\bs{\theta}}\left(\eta(\bs s,t),\eta(\bs s+\bs{h},t+u)\right)\right\},\label{boundaryterm_general2}
\end{align}
with 
\begin{align}\label{boundarypoints}
\mathcal{G}_{M,T}(\bs{h},u):=
\left\{(\bs s,t) \in \cals_M \times \mathcal{T}_T: \bs s + \bs h \notin \cals_M \textnormal{ or } t+u > T \right\}.
\end{align}
for $(\bs{h},u) \in \mathbb{N}^{d+1}$.
The pairwise maximum likelihood estimate (PMLE) is given by
\begin{align}
\wh{\bs{\theta}}=\arg\!\max_{\bs{\theta} \in \Theta}PL^{(M,T)}(\bs{\theta}). \label{PWLestimatesDef}
\end{align}
\end{definition}

We derive the asymptotic properties of the PMLE for two scenarios. The first one is based on regularly spaced observations with an increasing spatio-temporal domain. For this scenario we  follow the proofs in \citet{Steinkohl2} and show that the properties of strong consistency and asymptotic normality also hold if the dependence structure $\delta$ allows for spatially anisotropic effects as in \eqref{vario0}. In the second scenario, the observations are taken from a fixed spatial domain and an increasing temporal domain.

\subsection{Increasing spatio-temporal domain}\label{s32}

\ble\label{remG}
For $(\bs{h},u) \in \mathcal{H}_{\bs r} \times \{0,\ldots,p\}$, it holds that $$|\mathcal{G}_{M,T}(\bs{h},u)| \leq K_2( M^{d-1}{T}+{M^d)},$$ 
where $K_2$ is a constant independent of $M$ and $T$. 
\ele

\bproof
The number of space-time points within the space-time observation area, from which some grid point outside the observation area is within a lag $(\bs h, u) \in \mathcal{H}_{\bs r} \times \{1,\ldots,p\}$, is bounded by $M^{d-1}T\sum\limits_{j=1}^d r_j + M^dp$. 
 Thus we obtain
$$|\mathcal{G}_{M,T}(\bs{h},u)| \leq M^{d-1}T\sum_{j=1}^d r_j + M^dp \leq K_2 (M^{d-1}T+M^d),$$
where $K_2:=\max\big\{\sum_{j=1}^d r_j ,p\big\}$ is a constant independent of $M$ and $T$.
\eproof

\begin{theorem}[Strong consistency for large $M$ and $T$] \label{ThmConsistency} 
Let $\left\{\eta(\bs{s},t): \bs{s} \in \mathbb{R}^d, t \in [0,\infty) \right\}$ be a Brown-Resnick process as in \eqref{BR} with dependence structure
$$\delta(\bs{h},u) = \sum_{j=1}^d C_j |h_j|^{\alpha_j}+C_{d+1} |u|^{\alpha_{d+1}},\quad (\bs h,u)\in\R^{d+1},$$
where $0<\alpha_j\leq 2$ and $C_j> 0$ for $j=1,\ldots,d+1$.
Denote the parameter vector by
$$\bs{\theta}=(C_1,\ldots,C_{d+1}, \alpha_1, \ldots, \alpha_{d+1}).$$ 
 Assume that the true parameter vector $\bs{\theta}^{\star}$ lies in a compact set  
\begin{align}
\Theta^\star \subset \left\{(C_1,\ldots, C_{d+1}, \alpha_1, \ldots, \alpha_{d+1}): C_j \in (0, \infty), \alpha_j \in (0,2], \text{ } j=1, \ldots, d+1\right\}.\label{compactset}
\end{align} 
Suppose that the following identifiability condition holds for all $(\bs{s},t) \in \cals_M \times \mathcal{T}_T$:
\begin{align}
\bs{\theta}&=\tilde{\bs{\theta}} \quad\Leftrightarrow \label{identifiability}\\
 g_{\bs{\theta}}\left(\eta(\bs{s},t), \eta(\bs{s}+\bs{h},t+u)\right)&=g_{\tilde{\bs{\theta}}}\left(\eta(\bs{s},t), \eta(\bs{s}+\bs{h},t+u)\right), \quad \bs h \in \mathcal{H}_{\bs r}, \quad 0 \leq u \leq p\nonumber.
\end{align}\noindent
Then, the PMLE
$$\wh{\bs{\theta}}^{(M,T)}=\arg\!\max\limits_{\bs{\theta} \in \Theta^{\star}} PL^{(M,T)}(\bs{\theta})$$ 
is strongly consistent:
$$\wh{\bs{\theta}}^{(M,T)} \stas \bs{\theta}^{\star} \text{ as }M,T \rightarrow \infty.$$
\end{theorem}

\bproof
The proof uses the method of \citet{Wald}. One aim is to show that for some chosen maximum space-time lag $(\bs{r},p) \in \mathbb{N}_0^{d+1} \setminus\{\bs 0\}$ and $\bs{\theta} \in \Theta^\star$,
\begin{align*}
&\frac{1}{M^dT} PL^{(M,T)}(\bs{\theta}) \\
&=\frac{1}{M^dT}\Big({\sum\limits_{\bs s \in \mathcal{S}_M}} \sum\limits_{t=1}^T q_{\bs{\theta}}(\bs s,t;\bs{r},p) - \mathcal{R}^{(M,T)}(\bs{\theta})\Big) \stas \text{PL}(\bs{\theta}):=\mathbb{E}[q_{\bs{\theta}}(\bs{1},1;\bs{r},p)] 
\end{align*}
as $M,T \rightarrow \infty$. 
This is done by verifying the following two limit results:   Uniformly on $\Theta^{\star}$,
\begin{enumerate}[(A)]
\item \, 
$\dfrac{1}{M^dT}\sum\limits_{\bs s \in \mathcal{S}_M} \dsum\limits_{t=1}^T q_{\bs{\theta}}(\bs s,t;\bs{r},p) \stas \text{PL}(\bs{\theta}) \text{ as }M,T \rightarrow \infty,$
\item \,
$\dfrac{1}{M^dT} \mathcal{R}^{(M,T)}(\bs{\theta}) \stas 0 \text{ as }M,T \rightarrow \infty.$
\end{enumerate}
Furthermore, we need to show:
\begin{enumerate}[(A)]
\setcounter{enumi}{2}
\item \,
The limit function $\text{PL}(\bs{\theta})$ is uniquely maximized at the true parameter vector $\bs{\theta}^{\star} \in \Theta^{\star}$.
\end{enumerate}
We show (A). The almost sure convergence holds because $q_{\bs{\theta}}(\cdot)$ is a measurable function of lagged versions of $\eta(\bs{s},t)$ for $\bs{s} \in \cals_M$, $t \in \mathcal{T}_T$. 
Proposition~3 of \citet{Steinkohl2} implies a strong law of large numbers. 
What remains to show is that the convergence is uniform on the compact parameter space $\Theta^{\star}$. This can be done by carefully following the lines of the proof of Theorem~1 of \citet{Steinkohl2}, adapting it to the spatially anisotropic setting. For details we refer to \citet{Buhlma}, Theorem~4.4. We find that there is a positive finite constant $K_1$, independent of $\bs{\theta}, M$ and $T$, such that 
\begin{align}\label{K1}
\mathbb{E}\big[\big|\log g_{\bs{\theta}}\big(\eta(\bs{s}^{(1)},t^{(1)}),\eta(\bs{s}^{(2)},t^{(2)})\big)\big|\big] < K_1, \quad (\bs{s}^{(1)},t^{(1)}),(\bs{s}^{(2)},t^{(2)}) \in \mathbb{N}^{d+1},  
\end{align} 
and that $\mathbb{E}\big[\sup_{\bs{\theta} \in \Theta^{\star}}\left|q_{\bs{\theta}}(\bs{1},1;\bs{r},p)\right|\big] < \infty.$
Theorem~2.7 of \citet{Straumann1} implies that the convergence is uniform.\\[2mm]
Next we show (B). 
Using Proposition~3 of \citet{Steinkohl2} and (\ref{K1}) we have that, uniformly on $\Theta^{\star}$, 
\begin{align*}
&\sum\limits_{\bs{h} \in \mathcal{H}_{\bs{r}}} \sum\limits_{u=0}^p \frac{1}{|\mathcal{G}_{M,T}(\bs{h},u)|}\sum\limits_{(\bs s,t) \in \mathcal{G}_{M,T}(\bs{h},u)} \log\left\{g_{\bs{\theta}}\left(\eta(\bs s,t),\eta(\bs s+\bs{h},t+u)\right)\right\} \\ 
& \stas \,
\mathbb{E}\Big[\sum\limits_{\bs{h} \in \mathcal{H}_{\bs{r}}} \sum\limits_{u=0}^p \log\left\{g_{\bs{\theta}}\left(\eta(\bs{1},1),\eta(\bs{1}+\bs{h},1+u)\right)\right\}\Big] \text{ as } M,T \rightarrow \infty.
\end{align*}
 By Lemma~\ref{remG} and \eqref{K1} it follows that, uniformly on $\Theta^{\star}$,
\begin{align*}
&\frac{1}{M^dT}|\mathcal{R}^{(M,T)}(\bs{\theta})| \\ &\leq K_2{\big(\frac{1}{M}+\frac{1}{T}\big)} \bigg\vert{\sum\limits_{\bs{h} \in \mathcal{H}_{\bs{r}}} \sum\limits_{u=0}^p \frac{1}{|\mathcal{G}_{M,T}(\bs{h},u)|}\sum\limits_{(\bs s,t) \in \mathcal{G}_{M,T}(\bs{h},u)} \log\left\{g_{\bs{\theta}}\left(\eta(\bs s,t),\eta(\bs s+\bs{h},t+u)\right)\right\}}\bigg\vert \\ 
& \stas 0\quad\mbox{as $M,T \rightarrow \infty$, }
\end{align*}
Finally, we prove (C). Let $\bs{\theta} \neq \bs{\theta}^{\star}$. For $\bs s \in \cals_M$ and $t \in \mathcal{T}_T$, Jensen's inequality yields
\begin{align*}
\mathbb{E}\left[\log\left\{\frac{g_{\bs{\theta}}\left(\eta(\bs s,t),\eta(\bs s+\bs{h},t+u)\right)}{g_{\bs{\theta}^{\star}}\left(\eta(\bs s,t),\eta(\bs s+\bs{h},t+u)\right)}\right\}\right] 
&\leq \log \left\{\mathbb{E}\left[\frac{g_{\bs{\theta}}\left(\eta(\bs s,t),\eta(\bs s+\bs{h},t+u)\right)}{g_{\bs{\theta}^{\star}}\left(\eta(\bs s,t),\eta(\bs s+\bs{h},t+u)\right)}\right]\right\} \\
\lefteqn{\hspace*{-3cm}=\log \Big\{ \int\limits_{(0,\infty)^2}\frac{g_{\bs{\theta}}(y_1,y_2)}{g_{\bs{\theta}^{\star}}(y_1,y_2)}g_{\bs{\theta}^{\star}}(y_1,y_2) \,\mathrm{d}(y_1,y_2) \Big\}} \\
\lefteqn{\hspace*{-3cm} =\log \Big\{ \int\limits_{(0, \infty)^2} g_{\bs{\theta}}(y_1,y_2) \,\mathrm{d}(y_1,y_2) \Big\} \, = \, 0,}
\end{align*}
and it directly follows from \eqref{qteta} that $\text{PL}(\bs \theta) \leq \text{PL}(\bs \theta^{\star}).$ As $\bs \theta \neq \bs \theta^{\star}$, the identifiability condition (\ref{identifiability})  yields (C). \eproof

\brem
There are combinations of maximum space-time lags that lead to non-identifiable parameters, see Table~\ref{Table_Identifiability}. 
 However, Theorem~\ref{ThmConsistency} still applies to all identifiable parameters (cf. \citet{Steinkohl2}, Remark~2).  
\begin{table}[t] 
\centering
\begin{small}
\begin{tabular}{|p{0.7cm}|p{0.7cm}|p{0.7cm}|p{5cm}|} 
\hline
$r_1$ & $r_2$  & $p$  & \text{identifiable parameters}\\
\hline
1 & 0 & 0 & $C_1$\\
\hline
1 & 1 & 0 & $C_1$, $C_2$\\
\hline
1 & 1 & 1 & $C_1$, $C_2$, $C_3$\\
\hline
$>1$ & 0 & 0 & $C_1$, $\alpha_1$ \\
\hline
$>1$ & $>1$ & $>1$ & $C_1$, $\alpha_1$, $C_2$, $\alpha_2$, $C_3$, $\alpha_3$\\
\hline
\end{tabular}
\end{small}
\caption{Identifiable parameters for model \eqref{vario0} with $d=2$ for some examples of maximum space-time lags  $(r_1,r_2,p)$. }
\label{Table_Identifiability}
\end{table}
\erem

Next we prove asymptotic normality of the PMLE defined in (\ref{PWLestimatesDef}). As in the proof
of Theorem~\ref{ThmConsistency} we follow the lines of proof of \citet{Steinkohl2}, Section 5, adapting the arguments to the anisotropic setting. We start with some basic results needed throughout the remainder of the section. 

\begin{lemma}
\label{LemmaAsympnormalfirst}
Assume that all conditions of Theorem~\ref{ThmConsistency} are satisfied. Then for $\bs{s}^{(1)},\bs{s}^{(2)} \in \mathbb{R}^d$ and $t^{(1)}, t^{(2)} \in [0,\infty)$, the following assertions hold componentwise:
\begin{enumerate}[(1)]
\item 
The gradient of the bivariate log-density satisfies
$$\mathbb{E}\left[\left|\nabla_{\bs{\theta}} \log g_{\bs{\theta}}(\eta(\bs{s}^{(1)},t^{(1)}),\eta(\bs{s}^{(2)},t^{(2)}))\right|^3\right] < \infty, \quad \bs{\theta} \in \Theta^{\star}.$$
\item 
The Hessian matrix of the bivariate log-density satisfies
$$\mathbb{E}\left[\sup\limits_{\bs{\theta} \in \Theta^{\star}} \left|\nabla_{\bs{\theta}}^2 \log g_{\bs{\theta}} (\eta(\bs{s}^{(1)},t^{(1)}),\eta(\bs{s}^{(2)}, t^{(2)})) \right|\right] < \infty.$$
\end{enumerate}
\end{lemma}

\bproof
Assume identifiability of all parameters $C_j$, $\alpha_j$ for $j=1, \ldots, d+1$. 
For $y_1,y_2 \in (0, \infty)$ and for $(\bs{h},u) \in \mathbb{R}^{d+1}\setminus\{\bs 0\}$ lengthy but simple calculations of derivatives of \eqref{pairlike} yield
\begin{align*}
\nabla_{\bs{\theta}} \log g_{\bs{\theta}}(y_1,y_2)= \frac{\partial \log g_{\bs{\theta}}(y_1,y_2)}{\partial \delta(\bs{h},u)} \nabla_{\bs{\theta}}\delta(\bs{h},u),
\end{align*}
$$\frac{\partial \delta(\bs{h},u)}{\partial C_j}=|h_j|^{\alpha_j}, \quad \frac{\partial \delta(\bs{h},u)}{\partial \alpha_j}=C_j|h_j|^{\alpha_j} \log |h_j|,\quad j=1, \ldots d,$$
and 
$$\frac{\partial \delta(\bs{h},u)}{\partial C_{d+1}}=|u|^{\alpha_{d+1}}, \quad \frac{\partial \delta(\bs{h},u)}{\partial \alpha_{d+1}}=C_{d+1}|u|^{\alpha_{d+1}} \log |u|.$$
By compactness of the parameter space, as required in (\ref{compactset}), we can bound those first partial derivatives as well as the second order partial derivatives from above and below. So it remains to show that for $\bs{s}^{(1)},\bs{s}^{(2)} \in S$ and $t^{(1)},t^{(2)} \in T$,
$$\mathbb{E}_{\bs{\theta}^{\star}}\bigg[\bigg|\frac{\partial \log \{g_{\bs{\theta}}(\eta(\bs{s}^{(1)},t^{(1)}),\eta(\bs{s}^{(2)},t^{(2)}))\}}{\partial \delta(\bs{h},u)}\bigg|^3 \bigg] < \infty$$
and
$$\mathbb{E}_{\bs{\theta}^{\star}}\bigg[\sup\limits_{\bs{\theta} \in \Theta^{\star}} \bigg| \frac{\partial^2 \log \{g_{\bs{\theta}}(\eta(\bs{s}^{(1)},t^{(1)}),\eta(\bs{s}^{(2)},t^{(2)}))\}}{\partial^{2} \delta (\bs{h},u)} \bigg| \bigg] < \infty,$$
where the function $\delta(\bs{h},u)$ can be treated as a constant since it is bounded away from 0 by (\ref{compactset}). 
Hence, for the rest of the proof we refer to that of \citet{Steinkohl2}, Lemma~1. 
\eproof

For a central limit theorem we need certain mixing properties for a space-time setting (cf. \citet{Steinkohl2}, Section~5.1 and \citet{Huser}, Section~3.2).

\begin{definition}[Mixing coefficients and $\alpha$-mixing] \label{Defalphamixing}
Let $\{\eta(\bs{s},t): \bs s \in \mathbb{Z}^d,t \in \mathbb{N}\}$ be a space-time process. Let $d$ be some metric induced by a norm on $\mathbb{R}^{d+1}$.
For $\Lambda_1, \Lambda_2 \subset \mathbb{Z}^d \times \mathbb{N}$ let
$$d(\Lambda_1,\Lambda_2):=\inf\{d((\bs s^{(1)},t^{(1)}),(\bs s^{(2)},t^{(2)})): (\bs s^{(1)},t^{(1)}) \in \Lambda_1, (\bs s^{(2)},t^{(2)}) \in \Lambda_2\}.$$
\begin{enumerate}
\item [(1)]
For $k,\ell,n \geq 0$ the \textnormal{mixing coefficients} are defined as
\begin{align}\alpha_{k,\ell}(n)  := & \sup \{|\mathbb{P}(A_1 \cap A_2)-\mathbb{P}(A_1)\mathbb{P}(A_2)|: \notag \\ 
& A_1 \in \mathcal{F}_{\Lambda_1}, A_2 \in \mathcal{F}_{\Lambda_2}, |\Lambda_1| \leq k, |\Lambda_2| \leq \ell, d(\Lambda_1,\Lambda_2) \geq n \}, \label{alphamixingcoeff}
\end{align}
where $\mathcal{F}_{\Lambda_i}=\sigma(\eta(\bs s,t): (\bs s,t) \in \Lambda_i)$ for $i=1,2.$
\item[(2)] 
$\{\eta(\bs s,t): \bs s \in \mathbb{Z}^d, t \in \mathbb{N}\}$ is called $\alpha$\textnormal{-mixing} if for all $k,\ell>0$,
$$\alpha_{k,\ell}(n) \rightarrow 0, \quad n \rightarrow \infty.$$
\end{enumerate}
\end{definition}

Recall from Eq. (\ref{taildependencecoeffbrownresnick}) that for $(\bs{h},u) \in \mathbb{R}^{d+1}$ with $\delta$ as in \eqref{vario0} the tail dependence coefficient of the Brown-Resnick process is given by
$$\chi(\bs{h},u)=2\left(1-\Phi\left(\sqrt{\frac1{2}\big[C_1|h_1|^{\alpha_1}+ \cdots + C_d|h_d|^{\alpha_d}+C_{d+1}|u|^{\alpha_{d+1}}\big]}\right)\right).$$

Corollary~2.2 of \citet{Dombry} links the $\al$-mixing coefficients with the tail dependence coefficients, and we will use this for the next result.

\begin{proposition} \label{PropBolthausen_holds}
Let $\{\eta(\bs{s},t): \bs{s} \in \mathbb{R}^d, t \in [0,\infty)\}$ be the Brown-Resnick process \eqref{BR} with dependence function $\delta$ given by \eqref{vario0}. Then the process   $\{\eta(\bs s,t): \bs s \in \mathbb{Z}^d, t \in \mathbb{N}\}$ is $\alpha$-mixing, where the mixing coefficients in (\ref{alphamixingcoeff}) satisfy for $\mathcal{H}_{\bs{r}}$ as in \eqref{designmaskaniso}
\begin{enumerate}
\item[(1)]
$\sum\limits_{n=1}^{\infty} n^d \alpha_{k,\ell}(n) < \infty$ for $k+l \leq 4 (|\mathcal{H}_{\bs{r}}|+1)(p+1)$,
\item[(2)]
$\alpha_{(|\mathcal{H}_{\bs{r}}|+1)(p+1), \infty}(n)= o(n^{-(d+1)})$ as $\nto$,
\item[(3)]
$\sum\limits_{n=1}^{\infty} n^d \alpha_{(|\mathcal{H}_{\bs{r}}|+1)(p+1),(|\mathcal{H}_{\bs{r}}|+1)(p+1)}(n)^{\frac1{3}} < \infty.$
\end{enumerate}
\end{proposition}

\bproof
Note that for $(\bs{h},u) \in \mathbb{R}^{d+1}$, by the equivalence of norms, for some positive constant $L$,
\begin{align*}
d((\bs{h},u),(\bs{0},0))\leq \frac{1}{L} \max\{|h_1|,\ldots,|h_d|,|u|\}
\end{align*}
Therefore, for $n \in \mathbb{N}$, presuming $d((\bs{h},u),(\bs{0},0)) \geq n$ results in $\max\{|h_1|, \ldots, |h_d|, |u|\}$ $\geq Ln$, so that by Corollary~2.2 and Eq.~(3) of \citet{Dombry} we get
\begin{align}
\alpha_{k,\ell}(n) & \leq & 2k \ell \sup\limits_{d((\bs{h},u),(\bs{0},0)) \geq n} \chi(\bs{h},u)\, \leq \, 2k \ell \sup\limits_{\max\{|h_1|,\ldots, |h_d|, |u|\} \geq Ln}\chi(\bs{h},u),\label{alphabounds}\\
\alpha_{k, \infty} (n) &\leq & 2k \sum\limits_{d((\bs{h},u),(\bs{0},0)) \geq n}\chi(\bs{h},u) \, \leq \, 2k \sum\limits_{\max\{|h_1|,\ldots, |h_d|, |u|\} \geq Ln} \chi(\bs{h},u). 
\label{alphainftybounds}
\end{align}\noindent
In the following we use the notation $\Vert{(\bs h,u)}\Vert_{\infty}:=\max\{|h_1|,\ldots,|h_d|,|u|\}$ for $(\bs h,u) \in \mathbb{Z}^d \times \mathbb{N}.$
Using $1-\Phi(x) \leq \exp\{-\frac{1}{2}x^2\}$ for $x > 0$ and Eq.~\eqref{taildependencecoeffbrownresnick} and~\eqref{alphabounds}, we find for all $k, \ell \geq 0$,
\begin{align}
\alpha_{k,\ell}(n) &\leq 4k\ell \sup\limits_{\Vert{(\bs h,u)}\Vert_{\infty} \geq Ln} \left(1-\Phi(\sqrt{{\frac{\delta(\bs{h},u)}{2}}}\right) \notag \\
& \leq 4k\ell \sup\limits_{\Vert{(\bs h,u)}\Vert_{\infty} \geq Ln} \exp\left\{-{\frac{\delta(\bs{h},u)}{4}}\right\} \notag \\
&= 4k\ell \sup\limits_{\Vert{(\bs h,u)}\Vert_{\infty} \geq Ln} \exp\left\{-{\frac{1}{4}}\left[C_1|h_1|^{\alpha_1}+\ldots+C_d|h_d|^{\alpha_d}+C_{d+1}|u|^{\alpha_{d+1}}\right]\right\} \notag\\
& \leq 4k\ell \sup\limits_{\Vert{(\bs h,u)}\Vert_{\infty} \geq Ln} \exp\left\{-{\frac{1}{4}}\min\{|C_1|,\ldots,|C_{d+1}|\} \Vert{(\bs h,u)}\Vert_{\infty}^{\min\{\alpha_1,\ldots,\alpha_{d+1}\}}\right\} \notag\\
&\leq 4k\ell \exp\left\{-\frac{1}{4}\min\{|C_1|,\ldots,|C_{d+1}|\} (Ln)^{\min\{\alpha_1,\ldots,\alpha_{d+1}\}}\right\} \label{alphabound_fin}\\
&\rightarrow 0 \text{ as } n \rightarrow \infty. \notag
\end{align}
This implies $\al$-mixing.\\[2mm]
By similar arguments we obtain by \eqref{alphainftybounds} for all $k \geq 0$,
\begin{align}
&\alpha_{k,\infty}(n) \leq \notag \\
&4k\ \sum\limits_{\Vert{(\bs h,u)}\Vert_{\infty} \geq Ln} \exp\left\{-\frac{1}{4}\min\{|C_1|,\ldots,|C_{d+1}|\} \Vert{(\bs h,u)}\Vert_{\infty}^{\min\{\alpha_1,\ldots,\alpha_{d+1}\}}\right\}. \label{alphabound_inf}
\end{align}

We use the above bounds to prove assertions (1)-(3).\\
(1) \,  For $k+\ell \leq 4(|\mathcal{H}_{\bs{r}}|+1)(p+1)$ we have by {\eqref{alphabound_fin}},
\begin{align*}
&\sum\limits_{n=1}^{\infty} n^d \alpha_{k,\ell}(n)  
\leq 4k\ell \sum\limits_{n=1}^{\infty} n^d \exp\left\{-{\frac{1}{4}}\min\{|C_1|,\ldots,|C_{d+1}|\} {(Ln)}^{\min\{\alpha_1,\ldots,\alpha_{d+1}\}}\right\}  \,  < \, \infty.
\end{align*}
(2) \, First note that the number of grid points $(\bs h,u) \in \mathbb{R}^{d+1}$ with $\Vert{(\bs h,u)}\Vert_{\infty}=i$ for $i \in \mathbb{N}$ {equals $(i+1)^{d+1}-i^{d+1}$, and is therefore of order $\mathcal{O}(i^{d})$.} 
 We use {\eqref{alphabound_inf}} and a more precise estimate than in part (1) to obtain for sufficiently large $n$
\begin{align*}
& n^{d+1} \alpha_{(|\mathcal{H}_{\bs{r}}|+1)(p+1),\infty}(n) \\ 
& \leq 4n^{d+1} (|\mathcal{H}_{\bs{r}}|+1)(p+1) \\
& \quad \sum\limits_{\Vert{(\bs h,u)}\Vert_{\infty} \geq Ln} \exp\left\{-\frac{1}{4}\min\{|C_1|,\ldots,|C_{d+1}|\} \Vert{(\bs h,u)}\Vert_{\infty}^{\min\{\alpha_1,\ldots,\alpha_{d+1}\}}\right\} \\ 
& \leq K_3 n^{d+1} (|\mathcal{H}_{\bs{r}}|+1)(p+1) \sum\limits_{i = \lfloor{Ln\rfloor}}^{\infty} i^d \exp\Big\{-{\frac{1}{4}}\min\{C_1,\ldots,C_{d+1}\}i^{\min\{\alpha_1,\ldots,\alpha_{d+1}\}}\Big\}\\ & \rightarrow 0 \text{ as } n \rightarrow \infty,
\end{align*}
where $K_3$ 
is a positive constant. Convergence to 0 follows using the integral test for power series convergence and Lemma~\ref{aux}, Eq.~\eqref{formela1}.
\\
(3) \, We find, using again {\eqref{alphabound_fin}},
\begin{align*}
& \sum\limits_{n=1}^{\infty} n^d \alpha_{(|\mathcal{H}_{\bs{r}}|+1)(p+1),(|\mathcal{H}_{\bs{r}}|+1)(p+1)}(n)^{\frac{1}{3}}\\
 \leq &  \big(4\big[\,(|\mathcal{H}_{\bs{r}}|+1)(p+1)\big]^2\big)^{\frac{1}{3}} \\
& \cdot \sum\limits_{n=1}^{\infty} n^d \exp\Big\{-{\frac{1}{12}}\min\{C_1,\ldots,C_{d+1}\}(Ln)^{\min\{\alpha_1,\ldots,\alpha_{d+1}\}}\Big\}\\
< &  \infty
\end{align*}
 as in (1).
\eproof

Because of Lemma~\ref{LemmaAsympnormalfirst} and Proposition~\ref{PropBolthausen_holds} the following central limit theorem of \citet{Bolthausen} holds.

\begin{corollary} \label{CLTBolthausen}
Consider the process $\{\nabla_{\bs \theta} q_{\bs \theta^{\star}}(\bs s,t; \bs r, p): \bs s \in \mathbb{Z}^d, t \in \mathbb{N}\}$. Then 
\begin{align*}
\frac{1}{M^{\frac{d}{2}}\sqrt{T}}{\sum\limits_{\bs s \in \mathcal{S}_M}} \sum_{t=1}^T \nabla_{\bs{\theta}} q_{\bs{\theta}^{\star}}(\bs s,t;\bs{r},p) \std \mathcal{N}(\bs{0},\Sigma_1) \text{ as } M,T\to\infty,
\end{align*}
where
\begin{align}
\Sigma_1:=\sum\limits_{s_1=-\infty}^{\infty} \cdots \sum\limits_{s_d=-\infty}^{\infty} \sum\limits_{t=1}^{\infty} \textnormal{Cov}\left[\nabla_{\bs{\theta}}q_{\bs{\theta}^{\star}}(\bs{1},1; \bs{r},p),\nabla_{\bs{\theta}}q_{\bs{\theta}^{\star}}(s_1,\ldots,s_d,t;\bs{r},p)\right]. \label{Sigma1}
\end{align}
\end{corollary}

Now we formulate the main result of this section.

\begin{theorem}[Asymptotic normality for large $M$ and $T$] \label{ThmAsympNorm}
Assume the same conditions as in Theorem \ref{ThmConsistency}. 
Then
\begin{align}\label{AN}
\sqrt{M^dT}(\wh{\bs{\theta}}-\bs{\theta}^{\star}) \limd \mathcal{N}(\bs{0}, \tilde{\Sigma}_1) \mbox{ as } M,T \rightarrow \infty,
\end{align}
where $\tilde{\Sigma}_1:=F_1^{-1}\Sigma_1(F_1^{-1})^\top$ with $\Sigma_1$ given in \eqref{Sigma1} and
$$F_1:=\mathbb{E}\left[-\nabla_{\bs{\theta}}^2 q_{\bs{\theta}^{\star}}(\bs{1},1;\bs{r},p)\right].$$
\end{theorem}

\bproof
A Taylor expansion of the score function $\nabla_{\bs{\theta}}PL^{(M,T)}(\bs{\theta})$ around the true parameter vector $\bs{\theta}^{\star}$ yields for some
$\tilde{\bs{\theta}} \in [\wh{\bs{\theta}},\bs{\theta}^{\star}]:$
\begin{align*}
\bs{0}=\nabla_{\bs{\theta}}PL^{(M,T)}(\wh{\bs{\theta}})=\nabla_{\bs{\theta}}PL^{(M,T)}(\bs{\theta}^{\star})+\nabla_{\bs{\theta}}^2 PL^{(M,T)}(\tilde{\bs{\theta}})(\wh{\bs{\theta}}-\bs{\theta}^{\star}).
\end{align*}
Therefore,
\begin{align*}
M^{\frac{d}{2}}\sqrt{T}(\wh{\bs{\theta}}-\bs{\theta}^{\star})&=-\Big(\frac{1}{M^dT}\nabla_{\bs{\theta}}^2 PL^{(M,T)}(\tilde{\bs{\theta}})\Big)^{-1}\Big(\frac{1}{M^{\frac{d}{2}}\sqrt{T}}\nabla_{\bs{\theta}} PL^{(M,T)}(\bs{\theta}^{\star})\Big) \\
&=-\Big(\frac{1}{M^dT}{\sum\limits_{\bs s \in \mathcal{S}_M}} \sum\limits_{t=1}^{{T}} \nabla_{\bs{\theta}}^2 q_{\tilde{\bs{\theta}}}(\bs s,t;\bs{r},p) -\frac{1}{M^dT} \nabla_{\bs{\theta}}^2 \mathcal{R}^{(M,T)}(\tilde{\bs{\theta}})\Big)^{-1} \\ 
& \hspace{0.3cm}\Big(\frac{1}{M^{\frac{d}{2}}\sqrt{T}}{\sum\limits_{\bs s \in \mathcal{S}_M}} \sum\limits_{t=1}^{{T}} \nabla_{\bs{\theta}} q_{\bs{\theta}^{\star}}(\bs s,t;\bs{r},p) -\frac{1}{M^{\frac{d}{2}}\sqrt{T}} \nabla_{\bs{\theta}} \mathcal{R}^{(M,T)}(\bs{\theta}^{\star})\Big) \\
&=:-(I_1-I_2)^{-1}(J_1-J_2).
\end{align*}
Note the following:
\begin{itemize}
\item 
Corollary~\ref{CLTBolthausen} implies that $J_1\std \mathcal{N}(\bs{0},\Sigma_1)$ as $M,T\to\infty$. 
\item 
Using representation \eqref{boundaryterm_general2} of the boundary term $\mathcal{R}^{(M,T)}(\cdot)$ and Lemma~\ref{remG}, we find
\begin{align*}
&\Vert{J_2}\Vert=\frac{1}{M^{\frac{d}{2}}\sqrt{T}}\bigg\Vert{\sum\limits_{\bs h \in \mathcal{H}_{\bs r}} \sum\limits_{u=0}^p \sum\limits_{(\bs s,t) \in \mathcal{G}_{M,T}(\bs h,u)} \nabla_{\bs \theta} \log\{g_{\bs \theta^{\star}}(\eta(\bs s,t),\eta(\bs s+\bs h,t+u))\}}\bigg\Vert \\
& \leq \sqrt{K_2} \frac{\sqrt{M^{d-1}T+M^d}}{M^{\frac{d}{2}}\sqrt{T}}\\
&\hspace*{0.5cm}\bigg\Vert{\sum\limits_{\bs h \in \mathcal{H}_{\bs r}} \sum\limits_{u=0}^p  \frac{1}{\sqrt{|\mathcal{G}_{M,T}(\bs h,u)|}} \sum\limits_{(\bs s,t) \in \mathcal{G}_{M,T}(\bs h,u)} \nabla_{\bs \theta} \log\{g_{\bs \theta^{\star}}(\eta(\bs s,t),\eta(\bs s+\bs h,t+u))\}}\bigg\Vert \\
& \leq \sqrt{K_2}(\frac{1}{\sqrt{M}}+\frac{1}{\sqrt{T}}) \\
& \hspace*{0.5cm}\bigg\Vert{\sum\limits_{\bs h \in \mathcal{H}_{\bs r}} \sum\limits_{u=0}^p  \frac{1}{\sqrt{|\mathcal{G}_{M,T}(\bs h,u)|}} \sum\limits_{(\bs s,t) \in \mathcal{G}_{M,T}(\bs h,u)} \nabla_{\bs \theta} \log\{g_{\bs \theta^{\star}}(\eta(\bs s,t),\eta(\bs s+\bs h,t+u))\}}\bigg\Vert
\end{align*}
In the same way as done in Corollary~\ref{CLTBolthausen} for the process $\{\nabla_{\bs \theta} q_{\bs \theta^{\star}}(\bs s,t; \bs r, p): \bs s \in \mathbb{Z}^d, t \in \mathbb{N}\}$, we can apply Bolthausen's central limit theorem to the processes $\{\nabla_{\bs \theta} \log\{g_{\bs \theta^{\star}}(\eta(\bs s,t),\eta(\bs s + \bs h,t+u))\}: \bs s \in \mathbb{Z}^d, t \in \mathbb{N}\}$ for $\bs h \in \mathcal{H}_{\bs r}$, $u \in \{0,\ldots,p\}$. We conclude that
$$\sum\limits_{\bs h \in \mathcal{H}_{\bs r}} \sum\limits_{u=0}^p  \frac{1}{\sqrt{|\mathcal{G}_{M,T}(\bs h,u)|}} \sum\limits_{(\bs s,t) \in \mathcal{G}_{M,T}(\bs h,u)} \nabla_{\bs \theta} \log\{g_{\bs \theta^{\star}}(\eta(\bs s,t),\eta(\bs s+\bs h,t+u))\}$$ converges weakly to a normal distribution as $M,T\to\infty$, and it follows that $J_2\stp \bs{0}$ as $M,T\to\infty$.
\item 
As $\{\eta(\bs s,t): \bs s \in \mathbb{Z}^d, t \in \mathbb{N}\}$ 
is $\alpha$-mixing, the process
$$\{\nabla_{\bs{\theta}}^2q_{\bs{\theta}}(\bs s,t;\bs{r},p): \bs s \in \mathbb{Z}^d, t \in \mathbb{N}\}$$ is $\alpha$-mixing as a set of measurable functions of mixing lagged processes. 
Furthermore, as $\tilde{\bs{\theta}} \in [\wh{\bs{\theta}},\bs{\theta}^{\star}]$ and $\wh{\bs{\theta}}$ is strongly consistent, we have that $I_1\stas -F_1$ as $M,T\to\infty$. {The convergence is uniform on $\Theta^{\star}$} by Lemma~ \ref{LemmaAsympnormalfirst} which implies that $$\mathbb{E}\left[\sup\limits_{\bs{\theta} \in \Theta^{\star}} \left|\nabla_{\bs{\theta}}^2 q_{\bs{\theta}}(\bs{1},1;\bs{r},p)\right|\right] < \infty.$$
\item 
Concerning $I_2$, the law of large numbers applied to $$\left\{\nabla_{\bs{\theta}}^2 \log \{g_{\bs{\theta}}(\eta(\bs s,t),\eta(\bs s+\bs{h},t+u))\} : \bs s \in \mathbb{Z}^d, t \in \mathbb{N}\right\}$$ results in the fact that, in the same way as in part (B) of the proof of Theorem~\ref{ThmConsistency}, $I_2\stas \bs{0}$ as $M,T\to\infty$. 
\end{itemize}
Finally, summarising these results, Slutzky's Lemma yields \eqref{AN}.
\eproof

\subsection{Fixed spatial domain and increasing temporal domain}\label{s34}

As before we compute the PMLE based on observations on the area $\mathcal{S}_M \times \mathcal{T}_T$, but now we consider $M$ fixed, whereas $T$ tends to infinity.

We define the temporal $\alpha$-mixing coefficients (cf. \citet{Ibragimov}, Definition~17.2.1 or \citet{Bradley1}, Definition~1.6).

\begin{definition}[Temporal mixing coefficients and temporal $\al$-mixing]
Let $\{\eta(\bs s,t) : \bs s\in\cals_M, t\in\N\}$ be a space-time process.
Consider the metric $d(\cdot)$ of Definition~\ref{Defalphamixing}.
\begin{enumerate}[(1)]
\item 
Let  $\mathcal{T}^{(1)}, \mathcal{T}^{(2)} \subset \mathbb{N}$.
For $n\ge0$ the \textit{temporal $\alpha$-mixing coefficients} are defined as
\begin{align}\label{timemixingcoeff}
\al(n) := 
&\sup\{|P(A_1\cap A_2)-P(A_1)P(A_2)| : \nonumber\\
& A_1\in \calf_{\mathcal{S}_M\times \mathcal{T}^{(1)}}, A_2\in\calf_{\mathcal{S}_M\times \mathcal{T}^{(2)}}, d(\mathcal{S}_M\times \mathcal{T}^{(1)},\mathcal{S}_M\times \mathcal{T}^{(2)}) \geq n\},
\end{align} 
where $\calf_{\mathcal{S}_M\times \mathcal{T}^{(i)}}=\sigma(\eta(\bs s,t): (\bs s,t) \in \cals_M \times \mathcal{T}^{(i)})$ for $i=1,2.$
\item
$\{\eta(\bs s,t) : \bs s\in\cals_M, t\in\N\}$ is called \textit{ temporally $\al$-mixing}, if
\begin{align}\label{timemixing}
\al(n) \rightarrow 0, \quad n \rightarrow \infty.
\end{align}\noindent
\end{enumerate}
\end{definition}

\begin{proposition}\label{timemixingBR}
Let $\{\eta(\bs{s},t) : \bs{s}\in \mathbb{R}^d, t\in [0, \infty)\}$ be the Brown-Resnick process \eqref{BR} with dependence function $\delta$ given by \eqref{vario0}. Then the process $\{\eta(\bs{s},t) : \bs{s}\in\cals_M, t\in\N\}$ is temporally $\al$-mixing, where the mixing coefficients \eqref{timemixingcoeff} satisfy 
\begin{align}\label{cond2ANfixed}
\sum\limits_{n=1}^{\infty} |\alpha(n)|^{\frac{1}{3}} < \infty. 
\end{align}\noindent
\end{proposition}

\bproof
We use Eq.~(3) and Corollary~2.2 of \citet{Dombry} and \eqref{taildependencecoeffbrownresnick} to obtain for $n \in \mathbb{N}$
\begin{align*}
&\alpha(n) \\
&\leq 2 \sup_{d(\mathcal{S}_M \times \mathcal{T}^{(1)},\, \mathcal{S}_M\times \mathcal{T}^{(2)}) \geq n} \sum\limits_{(\bs{s}^{(1)},t^{(1)}) \atop \in \mathcal{S}_M\times \mathcal{T}^{(1)}} \sum\limits_{(\bs{s}^{(2)},t^{(2)}) \atop \in \mathcal{S}_M\times \mathcal{T}^{(2)}} \chi(\bs{s}^{(1)}-\bs{s}^{(2)},t^{(1)}-t^{(2)})  \\
&=4 \sup_{d(\mathcal{S}_M\times \mathcal{T}^{(1)}, \,\mathcal{S}_M\times \mathcal{T}^{(2)}) \geq n} \sum\limits_{(\bs{s}^{(1)},t^{(1)}) \atop \in \mathcal{S}_M\times \mathcal{T}^{(1)}}\sum\limits_{(\bs{s}^{(2)},t^{(2)}) \atop \in \mathcal{S}_M\times \mathcal{T}^{(2)}}\\ & \quad \bigg(1-\Phi\Big(\sqrt{\frac1{2}\big[C_1|s_1^{(1)}-s_1^{(2)}|^{\alpha_1}+ \cdots + C_d|s_d^{(1)}-s_d^{(2)}|^{\alpha_d}+C_{d+1}|t^{(1)}-t^{(2)}|^{\alpha_{d+1}}\big]}\Big)\bigg) \\
& \leq 4 M^{{2}d} \sup_{d(\mathcal{S}_M\times \mathcal{T}^{(1)}, \,\mathcal{S}_M\times \mathcal{T}^{(2)}) \geq n} \, \sum\limits_{(t^{(1)},t^{(2)}) \atop \in \mathcal{T}^{(1)} \times \mathcal{T}^{(2)}} 
\bigg(1-\Phi\Big(\sqrt{\frac1{2}\big[C_{d+1}|t^{(1)}-t^{(2)}|^{\alpha_{d+1}}\big]}\Big)\bigg)\\
&\leq 4 M^{{2}d} \sup_{d(\mathcal{S}_M\times \mathcal{T}^{(1)},\, \mathcal{S}_M\times \mathcal{T}^{(2)}) \geq n} \, \sum\limits_{(t^{(1)},t^{(2)}) \atop \in \mathcal{T}^{(1)} \times \mathcal{T}^{(2)}} 
\exp\Big\{-{\frac{1}{4}}C_{d+1} |t^{(1)}-t^{(2)}|^{\alpha_{d+1}}\Big\},
\end{align*}
where the last inequality follows from $1-\Phi(x) \leq \exp\{-\frac{1}{2}x^2\}$ for $x > 0$.
We bound $\alpha(n)$ for large $n$ further by
\begin{align*}
\alpha(n) \leq 4 M^{{2}d} \sum\limits_{t^{(1)} \in  \{-\infty, \ldots, 0\}} \sum\limits_{t^{(2)} \in \{n, \ldots, \infty\}} \exp\Big\{-{\frac{1}{4}}C_{d+1} |t^{(1)}-t^{(2)}|^{\alpha_{d+1}}\Big\}.
\end{align*}
In the double sum a temporal lag $u=|t^{(1)}-t^{(2)}| \geq n$ appears exactly $u-(n-1)$ times. This yields
\begin{align*}
\alpha(n) & \leq 4 M^{{2}d} \sum\limits_{u=n}^{\infty} (u-(n-1)) \exp\Big\{-{\frac{1}{4}}C_{d+1} u^{\alpha_{d+1}}\Big\} \\
& \leq 4 M^{{2}d} \sum\limits_{u=n}^{\infty} u \exp\Big\{-{\frac{1}{4}}C_{d+1} u^{\alpha_{d+1}}\Big\}.
\end{align*}
Convergence of the series \eqref{cond2ANfixed} now follows by the integral test and Lemma~\ref{aux}.
\eproof

In the following we show that strong consistency of the PMLE also holds, if the spatial domain remains fixed.

\begin{theorem}[Strong consistency for fixed $M$ and large $T$] \label{ThmConsistencySpacefix}
Assume the same conditions as in Theorem~\ref{ThmConsistency} restricted to the fixed space $\cals_M$.
Then the PMLE
$$\wh{\bs{\theta}}^{(M,T)}=\arg\!\max\limits_{\bs{\theta} \in \Theta^{\star}} PL^{(M,T)}(\bs{\theta})$$  
is strongly consistent, that is, $$\wh{\bs{\theta}}^{(M,T)} \stas \bs{\theta}^{\star} \text{ as } T \rightarrow \infty.$$
\end{theorem}

\bproof
For $\bs{\theta} \in \Theta^{\star}$ and $t \in \mathbb{N}$, set
$$q_{\bs{\theta}}^M(t;\bs r,p):={\sum\limits_{\bs s \in \mathcal{S}_M}}  \sum\limits_{\bs{h} \in \mathcal{H}_{\bs{r}} \atop \bs{s}+ \bs{h}\in\cals_M} \sum\limits_{u=0 \atop t+u \leq T}^p \mathbbmss{1}_{\{(\bs h, u) \neq (\bs 0,0)\}} \log \left\{g_{\bs{\theta}}\left(\eta(\bs{s},t), \eta(\bs{s}+\bs{h}, t+u) \right) \right\}.$$ Then 
\begin{align*}
&PL^{(M,T)}(\bs{\theta})=\sum\limits_{t=1}^T q_{\bs{\theta}}^M(t;\bs r,p).
\end{align*}
Following carefully the lines of the proof of Theorem \ref{ThmConsistency}, 
the following conditions hold for fixed spatial domain:
\begin{enumerate}[(A)]
\item 
$\dfrac{1}{T}\sum\limits_{t=1}^T q_{\bs{\theta}}^M(t;\bs{r},p) \stas PL^M(\bs{\theta}):=\mathbb{E}[(q_{\bs{\theta}}^M(1;\bs{r},p)]$
as $T \rightarrow \infty$ uniformly on the compact parameter space $\Theta^{\star}$.  
The main argument is that $q_{\bs{\theta}}^M(\cdot)$ is a function of temporally mixing lagged processes, then we apply again Theorem~2.7 of \citet{Straumann1}.
\item The limit function $PL^M(\bs{\theta})$ is uniquely maximised at the true parameter vector $\bs{\theta}^{\star} \in \Theta^{\star}$.
\end{enumerate}
\eproof

Now we formulate the main result of this section.

\begin{theorem}[Asymptotic normality for fixed $M$ and large $T$] \label{ThmANBRfixed}
Assume the same conditions as in Theorem~\ref{ThmConsistency} restricted to the fixed space $\cals_M$.
Then
\begin{align}\label{ANfixed}
\sqrt{T}(\wh{\bs{\theta}}-\bs{\theta}^{\star}) \limd \mathcal{N}(\bs{0}, \tilde{\Sigma}_2) \text{ as } T \rightarrow \infty,
\end{align}
where $\tilde{\Sigma}_2:=F_2^{-1}\Sigma_2(F_2^{-1})^\top$ with 
$$F_2:=\mathbb{E}[-\nabla_{\bs{\theta}}^2 q_{\bs {\theta}^\star}^M(1;\bs{r},p)]$$
and 
$$\Sigma_2:=\mathbb{V}\textnormal{ar}[\nabla_{\bs{\theta}} q_{\bs {\theta}^\star}^M(1;\bs{r},p)]+2\sum_{t=2}^{\infty} \mathbb{C}\textnormal{ov} [\nabla_{\bs \theta} q_{\bs{\theta}^{\star}}^M(1;\bs{r},p),\nabla_{\bs \theta} q_{\bs{\theta}^{\star}}^M(t;\bs{r},p)].$$
\end{theorem}

\bproof
By its definition as a function of lagged temporally mixing processes,  \linebreak
$(\nabla_{\bs \theta} q_{\bs{\theta}^{\star}}^M(t;\bs r,p))_{t \in \mathbb{N}}$  is also temporally $\alpha$-mixing with coefficients $\alpha'(n)=\alpha(n-p)$.
 Furthermore, 
\begin{align*}
\mathbb{E}\left[\nabla_{\bs \theta} \log \left\{g_{\bs{\theta^{\star}}}\left(\eta(\bs{0},0), \eta(\bs{h}, u) \right) \right\} \right]=0, \quad (\bs h ,u) \in \mathbb{N}_0^{d+1},
\end{align*}
because Lemma~\ref{LemmaAsympnormalfirst} implies regularity conditions of the pairwise log-likelihood \eqref{PWLgeneraldimension} allowing to interchange differentiation and integration.
Now note that Lemma~\ref{LemmaAsympnormalfirst} and Proposition~\ref{timemixingBR} imply that 
\begin{itemize}
\item 
$\mathbb{E} [|\nabla_{\bs{\theta}} q_{\bs{\theta}^{\star}}^M(t; \bs{r},p)|^{3}] < \infty$ for $t \in \mathbb{N}$ and every maximum spatial lag $\bs r$ and time lag $p$, and that
\item 
$\sum\limits_{n=1}^{\infty} |\alpha'(n)|^{\frac{1}{3}} < \infty.$
\end{itemize}
Therefore, the conditions of Theorem~18.5.3 of \citet{Ibragimov} (see also \citet{Bradley1}, Theorem~10.7) are satisfied and we conclude that
\begin{align}
\frac{1}{\sqrt{T}} \sum\limits_{t=1}^T \nabla_{\bs{\theta}} q_{\bs {\theta}^\star}^M(t;\bs{r},p) \limd \mathcal{N}(\bs{0}, \Sigma_2) \text{ as } T \rightarrow \infty. \label{CLTBradley}
\end{align}
Taylor expansion of the score function $\nabla_{\bs{\theta}}PL^{(M,T)}(\bs{\theta})$ around the true parameter vector $\bs{\theta}^{\star}$ yields for some
$\tilde{\bs{\theta}} \in [\wh{\bs{\theta}},\bs{\theta}^{\star}]:$
\begin{align*}
\bs{0}=\nabla_{\bs{\theta}}PL^{(M,T)}(\wh{\bs{\theta}})=\nabla_{\bs{\theta}}PL^{(M,T)}(\bs{\theta}^{\star})+\nabla_{\bs{\theta}}^2 PL^{(M,T)}(\tilde{\bs{\theta}})(\wh{\bs{\theta}}-\bs{\theta}^{\star}).
\end{align*}
Therefore,
\begin{align*}
\sqrt{T}(\wh{\bs{\theta}}-\bs{\theta}^{\star})&=-\Big(\frac{1}{T}\nabla_{\bs{\theta}}^2 PL^{(M,T)}(\tilde{\bs{\theta}})\Big)^{-1}\Big(\frac{1}{\sqrt{T}}\nabla_{\bs{\theta}} PL^{(M,T)}(\bs{\theta}^{\star})\Big) \\
&=-\Big(\frac{1}{T} \sum\limits_{t=1}^T \nabla_{\bs{\theta}}^2 q_{\tilde{\bs{\theta}}}^M(t;\bs{r},p) \Big)^{-1}\Big(\frac{1}{\sqrt{T}} \sum\limits_{t=1}^T \nabla_{\bs{\theta}} q_{\bs{\theta}^{\star}}^M(t;\bs{r},p) \Big) \, =: \, -I^{-1}J.
\end{align*}
Note the following:
\begin{itemize}
\item 
\eqref{CLTBradley} implies that $J\std \mathcal{N}(\bs{0},\Sigma_2)$ as $T\to\infty.$
\item 
Uniform convergence holds because of Lemma~\ref{LemmaAsympnormalfirst} which implies that componentwise 
$$\mathbb{E}\left[\sup\limits_{\bs{\theta} \in \Theta^{\star}} \left|\nabla_{\bs{\theta}}^2 q_{\bs{\theta}}^M(1;\bs{r},p)\right|\right] < \infty.$$
By temporal $\alpha$-mixing, since $\tilde{\bs{\theta}} \in [\wh{\bs{\theta}},\bs{\theta}^{\star}]$, and $\wh{\bs{\theta}}$ is strongly consistent, we have $I\stas-F_2$ as $T\to\infty$.
\end{itemize}
Finally, summarising those results, Slutzky's Lemma yields \eqref{ANfixed}.
\eproof

Throughout this section we have proved asymptotic properties of the parameter estimates of model \eqref{vario0} by classical results for ML estimators in combination with a spatio-temporal central limit theorem.
Such results can also be applied to other models like geometrically anisotropic models, provided the required rates for $\alpha$-mixing hold.

\section{Test for spatial isotropy}\label{s4}

We use the results of Section~\ref{s3} to formulate statistical tests for spatial isotropy versus anisotropy based on the model \eqref{vario0}, 
\beao
\delta(\bs{h},u) = \sum_{j=1}^d C_j |h_j|^{\alpha_j}+C_{d+1} |u|^{\alpha_{d+1}},
\eeao
for spatial lags $(\bs{h},u)=(h_1, \ldots, h_d,u) \in \mathbb{R}^{d+1}$.
We derive the necessary results for $d=2$. Generalisations to higher dimensions are possible, but notationally much more involved. Again we consider the two cases of an increasing and fixed spatial domain.

Due to the structure of model~\eqref{vario0} a test for isotropy versus anisotropy is a test of
\begin{align}\label{test}
H_0: \{C_1=C_2\,\mbox{ and } \, \al_1=\al_2\} \quad\mbox{versus}\quad
H_1: \{C_1\neq C_2\,\mbox{ or }\, \al_1\neq \al_2\}.
\end{align}\noindent

\subsection{Increasing spatial domain}\label{s41}

From Theorem~\ref{ThmAsympNorm} we know that, under suitable regularity conditions, the PMLE 
$$\wh{\bs{\theta}}=(\wh{C}_1,\wh{C}_2,\wh{C}_3,\wh{\alpha}_1, \wh{\alpha}_2, \wh{\alpha}_3)$$ 
is asymptotically normal; more precisely, for $M^2$ spatial observations on a regular grid and for $T$ equidistant time points we have
\begin{align}\label{asymptnorm_basis}
M \sqrt{T}\begin{pmatrix} \wh{C}_1-C_1  \\ \wh{C}_2-C_2 \\ \wh{C}_3-C_3 \\ \wh{\alpha}_1-\alpha_1 \\ \wh{\alpha}_2-\alpha_2 \\ \wh{\alpha}_3-\alpha_3 \end{pmatrix} \quad\limd \quad \mathcal{N}(\bs{0},\tilde{\Sigma}_1) \text{ as }  M,T \rightarrow \infty, 
\end{align}
where $\tilde{\Sigma}_1 \in \mathbb{R}^{6\times6}$ is the asymptotic covariance matrix given in  Theorem~\ref{ThmAsympNorm}. 

Our test is based on the spatial parameters only. Moreover, we test the two equalities in $H_0$ separately and use Bonferroni's inequality to solve the multiple test problem.

\ble \label{asymp_distr}
Assume the conditions of Theorem~\ref{ThmAsympNorm}. Setting $A_1:=(-1,1,0,0,0,0)$ and $A_2:=(0,0,0,-1,1,0)$, we have that, as $M,T\to\infty$,
\begin{align}
M\,\sqrt{T}((\wh{C}_2-\wh{C}_1)-(C_2-C_1)) & \limd \mathcal{N}(0,A_1\tilde{\Sigma}_1A_1^\top), \label{limitlaw1}\\
M\,\sqrt{T}((\wh{\alpha}_2-\wh{\alpha}_1)-(\alpha_2-\alpha_1)) & \limd \mathcal{N}(0,A_2\tilde{\Sigma}_1A_2^\top). \label{limitlaw2}
\end{align}\noindent
\ele

\bproof 
We obtain the left hand side of (\ref{limitlaw1}) and (\ref{limitlaw2})  by multiplying $A_1$ and $A_2$ to \eqref{asymptnorm_basis}, respectively.
This yields the limits on the right hand side by the continuous mapping theorem.
\eproof

We define 
$$\theta_C:=(C_2-C_1),\, \wh{\theta}_C:=(\wh{C}_2-\wh{C}_1),\quad \theta_{\alpha}:=(\alpha_2-\alpha_1),\, \wh{\theta}_{\alpha}:=(\wh{\alpha}_2-\wh{\alpha}_1).$$ 
Then the multiple test problem \eqref{test} becomes
\begin{eqnarray}
\label{test0}
H_{0,1}: \{\theta_C=0\} & \mbox{versus} & H_{1,1}: \{\theta_C\neq 0\}\\
H_{0,2}: \{\theta_\al=0\} & \mbox{versus} & H_{1,2}: \{\theta_\al\neq 0\}.
\end{eqnarray}\noindent
Since the variances in \eqref{limitlaw1} and \eqref{limitlaw2} are not known explicitly, we find the rejection areas of the two tests by subsampling as suggested in \citet{Politis4}, Chapter~5. Their main Assumption~5.3.1, the existence of a weak limit law of the estimates, is satisfied by Lemma~\ref{asymp_distr}.

We formulate the subsampling procedure in the terminology of the space-time process $\{\eta(\bs{s},t): \bs{s} \in \cals_M, t \in \mathcal{T}_T\}.$
We choose space-time block lengths \linebreak
$\bs{b}=(b_1,b_2, b_3)\ge (1,1,1) $ and the degree of overlap $\bs{e}=(e_1,e_2, e_3) \leq (M,M,{{T}})$. 
The blocks are indexed by $\bs i=(i_1,i_2,i_3) \in \mathbb{N}^3$ with $i_j \leq q_j$ for $q_j:=\lfloor \frac{M-b_j}{e_j}\rfloor + 1$, $j=1,2$ and $q_3:=\lfloor \frac{T-b_j}{e_j}\rfloor + 1.$ This results in a total number of $q= q_1 q_2 q_3$ blocks, which we summarise in the set
\beao 
E_{\bs{i},\bs{b},\bs{e}}
&=& \big\{(s_1, s_2,t) \in \cals_M \times \mathcal{T}_T: (i_j-1)e_j+1 \leq s_j \leq (i_j-1)e_j+b_j, j=1,2, \\
&& \quad(i_{3}-1)e_{3}+1 \leq t \leq (i_{3}-1)e_{3}+b_{3}\big\}.
\eeao
Now we estimate $\theta_C$ and $\theta_\al$ based on all observations in a block, hence getting $q$ different estimates, which we denote by
$\wh\theta_{C,{\bs b}, \bs{i}}$ and $\wh\theta_{\al,{\bs b}, \bs{i}}$.

In order to find rejection areas for the isotropy test, we will use Lemma~\ref{asymp_distr}, and take care of the unknown variance in the normal limit by a subsampling result.

\begin{theorem}\label{subsampling_conf}
Denote by $\tau_{M,T}:= M \sqrt{T}$ and $\tau_{\bs b} =\sqrt{b_1 b_2 b_{3}}$ the square roots of the number of observations in total and in each block, respectively.
Assume that the conditions of Theorem~\ref{ThmAsympNorm} hold and, as $M,T \rightarrow \infty$,  
\begin{enumerate}[(i)]  
\item $b_i \rightarrow \infty$ for  $i=1,2,3,$ such that
${b_i}=o({M})$  for  $i=1,2,$ and $b_{3} = o(T)$ (hence, $\tau_{\bs{b}}/\tau_{M,T} \rightarrow 0$),
\item $\bs{e}$ does not depend on $M$ or $T$.
\end{enumerate}
In the following $\wh\theta$ stands for either $\wh\theta_C$ or $\wh\theta_\al$.
Define the empirical distribution function
\begin{align}
L_{\bs{b},\wh\theta}(x):=\frac{1}{q} \sum\limits_{i_1=1}^{q_1} \sum\limits_{i_2=1}^{q_2} \sum\limits_{i_{{3}}=1}^{q_{{3}}} \bs{1}_{\left\{\tau_{\bs{b}}\left|\wh{\theta}_{\bs{b}, \bs{i}}-\wh{\theta}\right| \leq x\right\}},\quad x\in\R,
\end{align}
and the empirical quantile function
\begin{align} 
c_{\bs{b},\wh\theta}(1-\beta):=\inf\left\{x\in\R :L_{\bs{b},\wh\theta}(x) \geq 1-\beta\right\},\quad \beta\in(0,1).
\end{align} 
Then the following statements hold for $M, T \rightarrow \infty$:
\begin{enumerate}
\item[(1)]  \label{subs1}
Denote by $\Phi_{\sigma}(\cdot)$ the distribution function of a mean 0 normal random variable $Z$
with variance 
$$\sigma^2=\begin{cases} A_1\tilde{\Sigma}_1A_1^\top, \quad\text{ in case of }\, \wh\theta_C,\\ A_2\tilde{\Sigma}_1A_2^\top, \quad\text{ in case of }\, \wh\theta_\alpha, \end{cases}$$
and recall that $2\Phi_{\sigma}(\cdot)-1$ is the distribution function of $|Z|$.
Then
$$L_{\bs{b},\wh\theta}(x) \quad\stp\quad 2\Phi_{\sigma}(x)-1, \quad x \in \mathbb{R}.$$
\item[(2)] \label{subs2}
Set $J_{\wh\theta}(x):=\mathbb{P}(\tau_{M,T}|\wh{\theta}-\theta| \leq x )$ for $x\in\R$, then
$$\sup\limits_{x \in \mathbb{R}}
\left|L_{\bs{b},\wh\theta}(x)-J_{\wh\theta}(x)\right| \quad\stp\quad  0.$$
\item[(3)]  \label{subs3}
For $\beta\in(0,1)$, 
\begin{align}
\mathbb{P}\left(\tau_{M,T}|\wh{\theta}-\theta| \leq c_{\bs{b},\wh\theta}(1-\beta) \right) \rightarrow 1-\beta. \label{Motiv_crit_value_abs}
\end{align}
\end{enumerate} 
\end{theorem}

\bproof
We apply Corollary~5.3.1 of \citet{Politis4}. Their main Assumption~5.3.1; i.e., the existence of a continuous limit distribution, is satisfied by Lemma~\ref{asymp_distr}. Assumptions (i)-(ii) are also presumed by \citet{Politis4}. 
The required condition on the $\alpha$-mixing coefficients is satisfied similarly as in the proof of Proposition~\ref{PropBolthausen_holds} by Lemma~\ref{aux} and the result holds.
\eproof 

From (\ref{Motiv_crit_value_abs}), we find rejection areas for the test statistics $\tau_{M,T}\wh{\theta}$ at confidence level $\beta\in(0,1)$ as (recall that $\wh\theta$ stands for either $\wh\theta_C$ or $\wh\theta_{\alpha}$)
$$\textnormal{Rej}_{\wh\theta}^{(M,T)}:=(-\infty, -c_{\bs{b},\wh\theta}(1-\beta)) \cup (c_{\bs{b},\wh\theta}(1-\beta), \infty)=[-c_{\bs{b},\wh\theta}(1-\beta),c_{\bs{b},\wh\theta}(1-\beta)]^c.$$
Bonferroni's inequality 
$$\mathbb{P}(\mbox{reject $H_{0,1}$ or $H_{0,2}$}) \leq
 \mathbb{P}(\mbox{reject $H_{0,1}$}) + \mathbb{P}(\mbox{reject $H_{0,2}$}) \leq 2 {\beta},$$
applies and solves the multiple test problem.

\subsection{Fixed spatial domain}\label{s42}

First note that an analogue of Lemma~\ref{asymp_distr} holds with rate $\sqrt{T}$ instead of $M\sqrt{T}$ and with the asymptotic covariance matrix $\tilde{\Sigma}_2$ as given in Theorem~\ref{ThmANBRfixed}.

The subsampling statement corresponding to Theorem \ref{subsampling_conf} then reads as follows.

\begin{theorem}\label{subsampling_confFixed}
Denote by $\tau_{T}:= \sqrt{T}$ and $\tau_{b_3} =\sqrt{b_3}$ the square roots of the number of time points of observations in total and in each block, respectively.
Assume that the conditions of Theorem~\ref{ThmANBRfixed} are satisfied and that Lemma~\ref{asymp_distr} holds for $T\to\infty$ with rate $\sqrt{T}$ instead of $M\sqrt{T}$ and with the asymptotic covariance matrix $\tilde{\Sigma}_2$ as given in Theorem~\ref{ThmANBRfixed}. 
Assume further that as $T \rightarrow \infty$, 
\begin{enumerate}[(i)]
\item $b_3 \rightarrow \infty$ such that
$b_{3} = o(T)$ (hence, $\tau_{b_3}/\tau_{T} \rightarrow 0$),\label{subfix1}
\item $\bs{e}$ does not depend on $T$,
\item $b_1, b_2 \rightarrow M$. \label{subfixlast}
\end{enumerate}
Let $\bs b=(b_1,b_2,b_3)$, $\tau_{\bs b}=\sqrt{b_1b_2b_3}$ and $\tau_{M,T}=M\sqrt{T}$. 
With $\tilde{\Sigma}_1$ as in Theorem~\ref{subsampling_conf} replaced by $M^2\tilde{\Sigma}_2$, conclusions (a), (b), and (c) of Theorem~\ref{subsampling_conf} remain true as $T$ tends to infinity.  
\end{theorem}

\bproof
We apply Corollary~5.3.2 of \citet{Politis4}. The required temporal mixing condition is satisfied  similarly as in the proof of Proposition~\ref{timemixingBR} by Lemma~\ref{aux}. 
\eproof

\begin{remark}\rm
We can in practice apply the same procedure of subsampling as in Section~\ref{s41}. 
This is justified by the fact that ${\tau_{b_3}}/{\tau_{T}} \rightarrow 0$ implies that ${\tau_{\bs b}}/{\tau_{M,T}} \rightarrow 0$ as $T \rightarrow \infty$ under conditions~(\ref{subfix1})-(\ref{subfixlast}) of Theorem~\ref{subsampling_confFixed}. 
In particular, the rejection area for $\tau_{T}\wh\theta$ (where again $\wh\theta$ stands for either $\wh\theta_C$ or $\wh\theta_{\alpha}$) is found as
$$\textnormal{Rej}_{\wh\theta}^{(T)}:=\frac1{M}{\textnormal{Rej}_{\wh\theta}^{(M,T)}}.$$
\end{remark}

\section{Data analysis}\label{s5}

\begin{figure}[t] 
\centering
\includegraphics[width=12cm,height=9cm]{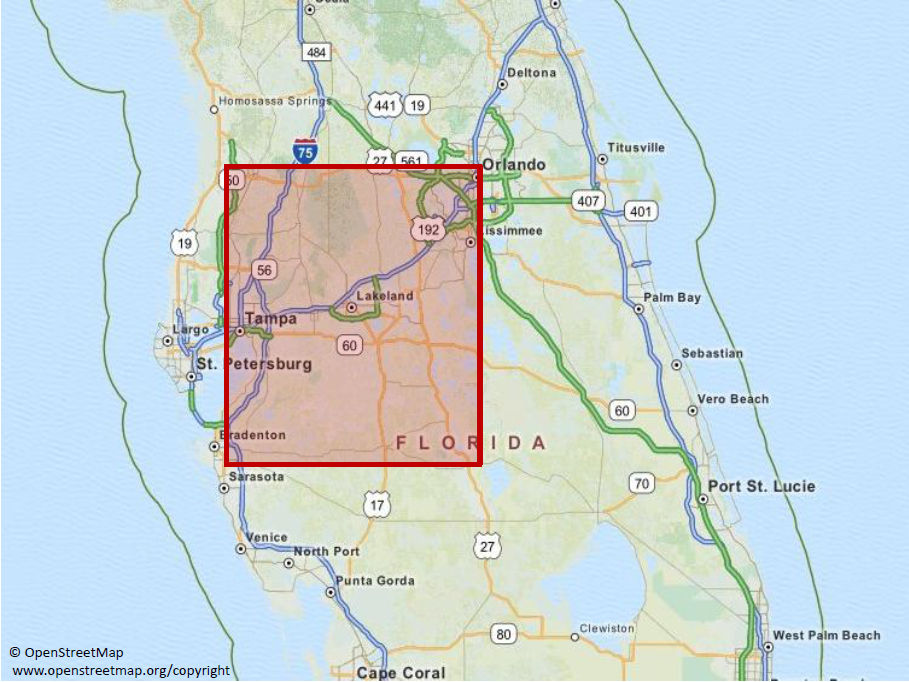}
\caption{Rainfall observation area in Florida}
\label{Florida}
\end{figure}

We fit the Brown-Resnick space-time process \eqref{BR} with dependence structure  given by the model \eqref{vario0}
to radar rainfall data, which were provided by the Southwest Florida Water Management District (SWFWMD).
The data used for the analysis are rainfall measurements on a square of 120km$\times$120km in Florida (see Figure~\ref{Florida}) over the years 1999-2004.
The raw data consist of measurements in inches on a regular grid in space every two kilometres and every 15 minutes.
Since there exist wet seasons and dry seasons with almost no rain we consider only the wet season June-September. 
Moreover, the area is basically flat with predominant easterly winds due to its closeness to the equator and,  therefore, existing trade winds.
Hence, \eqref{vario0} with parameters that possibly differ along both spatial axes fits well without introducing a rotation matrix.

\subsection{Data transformation and marginal modelling} \label{s51}

We carry out a block-maxima method in space and time as follows: 
We calculate cumulated hourly rainfall by adding up four consecutive measurements. Then we take block-maxima over 24 consecutive hours and over 10km$\times$10km areas; i.e., the daily maxima over 25 locations, resulting in a $12\times12$ grid in space for all $6\times 122$ days of the wet seasons giving a time series of dimension $12 \times 12$ and of length 732. 
Taking smaller areas than 10km$\times$10km squares or a higher temporal resolution (e.g. 12-hour-maxima) results in observations that are not max-stable and the max-stability test described in Section~\ref{s52} would reject.

By removing possible seasonal effects, we transform the data to stationarity.
We obtain the observations
\begin{align}
\left\{\tilde{\eta}((s_1,s_2),t): s_1,s_2=1,\ldots,12, t=1,\ldots,732 \right\}. \label{DataGumbel1}
\end{align}
Taking daily maxima removes for every location most of the dependence in the time series. This implies that marginal parameter estimates found by maximum likelihood estimation are consistent and asymptotically normal.

To give some details: 
for each fixed location $(s_1,s_2)$, we fit a univariate \textit{generalised extreme value distribution} (cf. \citet{Embrechts}, Definition~3.4.1) to the associated time series. The estimated shape parameters are all sufficiently close to 0 to motivate a Gumbel distribution as appropriate model. We therefore fit a Gumbel distribution $\Lambda_{\mu,\sigma}(x)=\exp\{-\text{e}^{-\frac{x-\mu}{\sigma}}\}$ with parameters $\mu=\mu(s_1,s_2) \in \mathbb{R}$ and $\sigma=\sigma(s_1,s_2) > 0$ and obtain estimates $\wh{\mu}=\wh{\mu}(s_1,s_2)$ and $\wh{\sigma}=\wh{\sigma}(s_1,s_2)$.

Depending on different statistical questions and methods, we transform \eqref{DataGumbel1} either to standard Gumbel or standard Fr\'echet margins. In the first case we set 
\begin{align}
\eta_1((s_1,s_2),t):=\frac{\tilde{\eta}((s_1,s_2),t)-\wh{\mu}}{\wh{\sigma}}, \quad t=1,\ldots,732, \label{DataGumbel2}
\end{align}
and in the latter case, with $\Lambda_{\wh{\mu},\wh{\sigma}}$ denoting the Gumbel distribution with estimated parameters,
\begin{align}
\eta_2((s_1,s_2),t):=-\frac{1}{\log\left\{\Lambda_{\wh{\mu},\wh{\sigma}}(\tilde{\eta}((s_1,s_2),t))\right\}}, \quad t=1,\ldots,732. \label{DataFrechet}
\end{align}

\begin{figure}[t] 
\centering
\subfloat[]{\includegraphics[height=6cm,width=6.92cm]{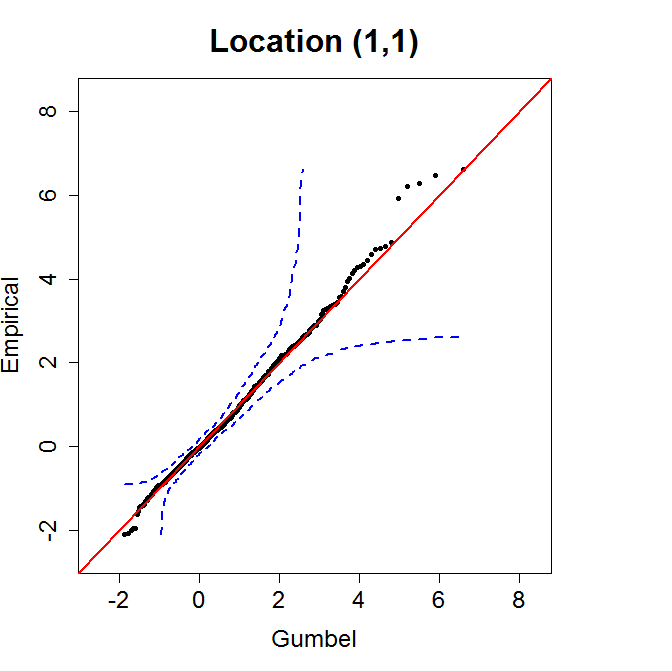}}
\subfloat[]{\includegraphics[height=6cm,width=6.92cm]{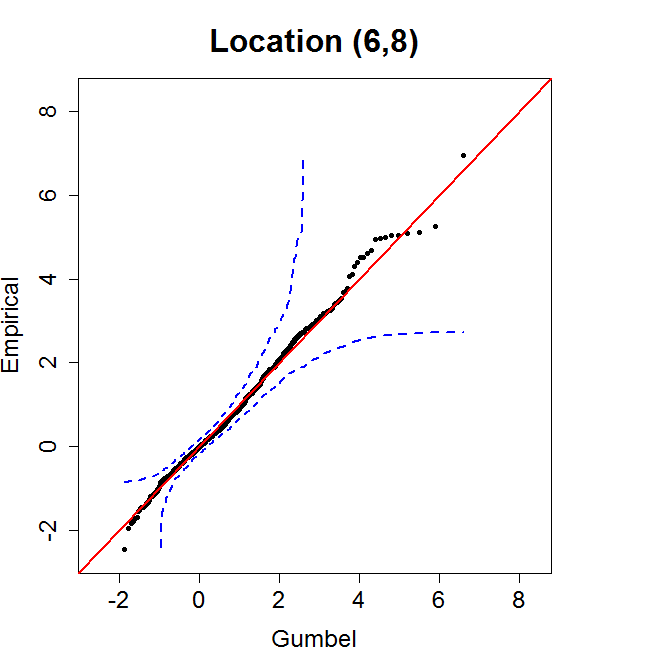}} \\
\subfloat[]{\includegraphics[height=6cm,width=6.92cm]{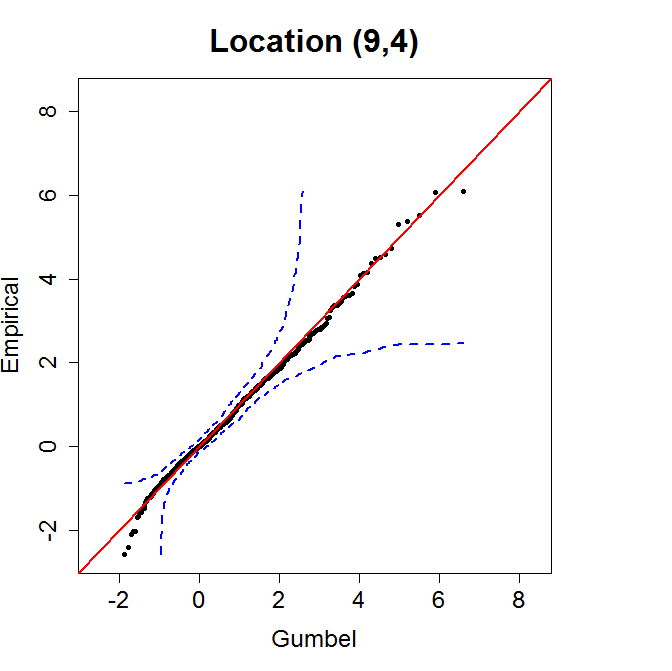}} 
\subfloat[]{\includegraphics[height=6cm,width=6.92cm]{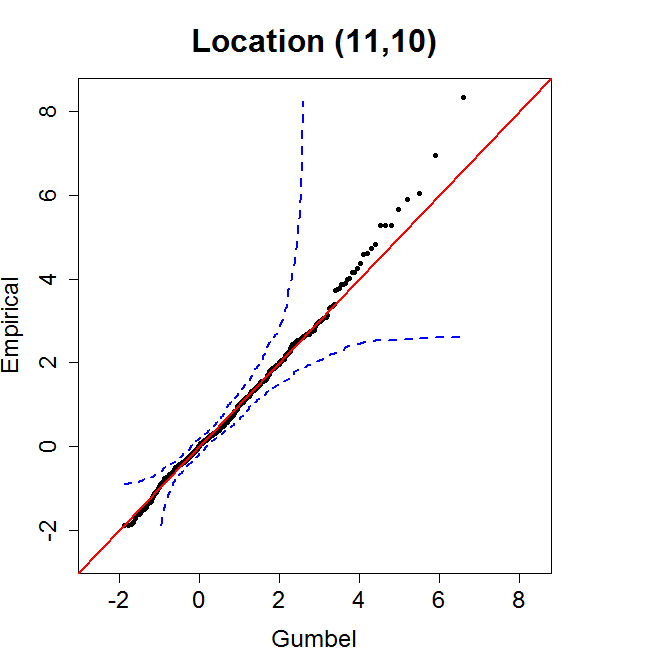}}
\caption{qq-plots of the Gumbel transformed time series values versus the standard Gumbel distribution for four locations: (1,1) (top left), (6,8) (top right), (9,4) (bottom left) and (11,10) (bottom right). Dashed blue lines mark $95 \%$ confidence bounds. Solid red lines correspond to no deviation.}
\label{qq_margins}
\end{figure}

We assess the goodness of the marginal fits by qq-plots of the observations \eqref{DataGumbel2} versus the standard Gumbel quantiles for every spatial location.
Figure~\ref{qq_margins} depicts the qq-plots at four exemplary spatial locations $(1,1)$, $(6,8)$, $(9,4)$ and $(11,10)$. \footnote{We use the \texttt{R}-package \texttt{extRemes} (\citet{extRemes}).}
 Confidence bounds are based on the Kolmogorov-Smirnov statistic (cf. \citet{Doksum}, Theorem~1 and Remark~1). 
All graphs show a reasonably good fit.

In the following data analysis we regard \eqref{DataFrechet} as realisations of the space-time Brown-Resnick process \eqref{BR} with dependence structure $\delta$ as in \eqref{vario0}:
\begin{align}\label{3D}
\delta(h_1,h_2,u)=C_1|h_1|^{\alpha_1}+C_2|h_2|^{\alpha_2}+C_3|u|^{\alpha_3},
\end{align}\noindent 
with $h_1=s_1^{(1)}-s_1^{(2)}$, $h_2=s_2^{(1)}-s_2^{(2)}$, $u=t^{(1)}-t^{(2)}$, for two spatial locations $\bs{s}^{(1)}=(s_1^{(1)},s_2^{(1)})$ and $\bs{s}^{(2)}=(s_1^{(2)},s_2^{(2)})$ and two time points $t^{(1)}$ and $t^{(2)}$. 

\subsection{Testing for max-stability in the data}\label{s52}

We first want to check if the block-maxima data originate from a max-stable process. 
A diagnostic tool is based on a multivariate Gumbel model (cf. \citet{Wadsworth}), and we explain first the method in general.
We assume a space-time model of a general spatial dimension $d \in \mathbb{N}$.
As before, we denote the regular grid of space-time observations by
$$\mathcal{S}_M\times \mathcal{T}_T=\{1,\ldots,M\}^d \times \{1, \ldots, T\}.$$
We define a hypothesis test based on the standard Gumbel transformed space-time observations \eqref{DataGumbel2} by
\begin{align}\label{testms}
H_0: \{\eta_1(\bs{s},t): (\bs{s},t) \in \mathbb{R}^d \times [0, \infty)\} \textit{ is max-stable. }
\end{align}\noindent 
Under $H_0$ all finite-dimensional margins are max-stable; particularly, for every $D \subseteq \mathcal{S}_M\times \mathcal{T}_T$, the multivariate distribution function of  $\{\eta_1(\bs{s},t): (\bs{s},t) \in D\}$ is given by
$$G_D(y_1,\ldots,y_{|D|})=\exp\{-V_D(\textnormal{e}^{y_1},\ldots,\textnormal{e}^{y_{|D|}})\}, \quad (y_1,\ldots,y_{|D|}) \in \mathbb{R}^{|D|},$$
where $V_D$ is the exponent measure from \eqref{exponentmeasureD}.
Since $V_D$ is homogeneous of order -1, the random variable $$\eta_D:=\max\{\eta_1(\bs{s},t):(\bs{s},t) \in D\}$$
has univariate Gumbel distribution function 
\begin{align}\label{etaD}
\mathbb{P}(\eta_D \leq y)=G_D(y,\ldots,y)
=\exp\{-\textnormal{e}^{-y}V_D(1,\ldots,1)\}=\textnormal{e}^{-\textnormal{e}^{-(y-\mu_D)}},\quad y\in\R;
\end{align}
i.e., $\mu_D:=\log V_D(1,\ldots,1)$ is the location parameter and, since $1 \leq V_D(1,\ldots,1) \leq |D|$, we have $0 \leq \mu_D \leq \log|D|$. 
These considerations can be used to construct a graphical test for max-stability: 
First, choose different subsets $D$ with the same fixed cardinality. Then extract several independent realisations of the random variables $\eta_D$ from the data and test by means of a qq-plot, if they follow a Gumbel distribution.

We apply this test to the standardized Gumbel transformed data \eqref{DataGumbel2}.
As indicated above, taking daily maxima removes for every location most of the dependence in the time series. For this test we want to take every precaution to make sure that we work indeed with independent data.
Preliminary tests show that spatial observations, which are a small number of $B_2$ days apart (to be specified below), show only very little time-dependence.
 
\begin{figure}[t]
\centering
\includegraphics[width=\linewidth]{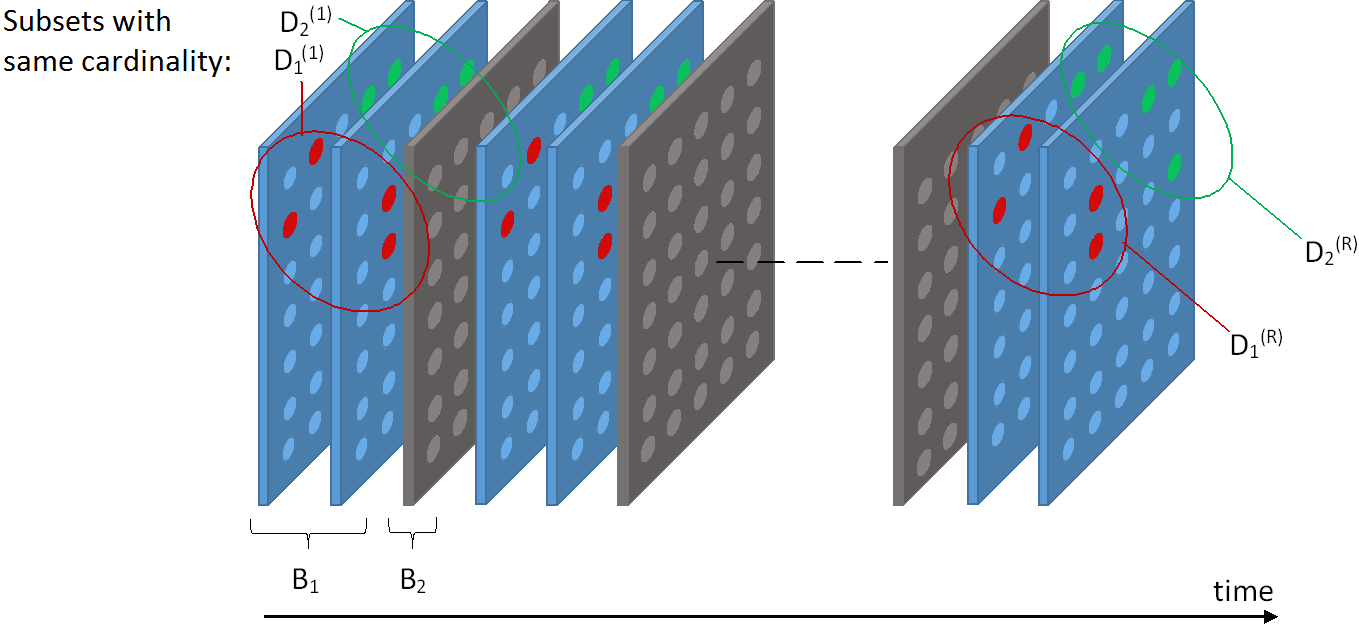}
\caption{$R$ independent realisations of $\eta_D$ for different subsets $D$ of the space-time observation area.}
\label{blocks_ill}
\end{figure}

Consequently, we define time blocks of size $B_1$ of spatial observations, which are  in turn separated by time blocks of size $B_2$ as
\begin{align}\label{blocks}
\mathcal{S}_M\times \mathcal{T}^{(i)}= \{1,\ldots,M\}^2 \times \{(i-1)(B_1+B_2)+t: \quad t=1,\ldots,B_1\}, 
\end{align}
for $i=1, \ldots, R=\lfloor{\frac{T}{B_1+B_2}}\rfloor.$ The numbers $B_1$ and $B_2$ need to be chosen in such a way that the blocks can be considered as independent.
This results in $R$ independent time blocks of length $B_1$ of spatial data and thus in $R$ independent realisations of $\eta_D$ for every $D \subseteq \mathcal{S}_M\times \{1, \ldots, B_1\}.$ 
The procedure is illustrated in Figure~\ref{blocks_ill}.

We use these i.i.d. realisations to estimate $\mu_D$ for every $D$ by maximum likelihood estimation restricted to $[0,\log |D|]$. 
Since the MLE of the location parameter of a Gumbel distribution is not unbiased (cf. \citet{Kotz}, Section~9.6), we perform a bias correction. 

For the diagnostic we take $K \in \mathbb{N}$ and consider subsets $D$ with cardinality $|D|=K$.  As the total number $\binom{B_1 M^2}{K}$ of those subsets is in most cases intractably large, we randomly choose $m:=\min\{R,\binom{B_1 M^2}{K}\}$ subsets and obtain in total $N= m\cdot R$ subsets, which we denote by $D_j^{(i)}$ for $j=1, \ldots, m$ and $i=1,\ldots,R$.
For every $j=1,\ldots,m$ we estimate $\mu_{D_j}$ by MLE based on the i.i.d. random variables $\eta_{D_j}^{(i)}:=\eta_{D_j^{(i)}}$, $i=1,\ldots,R.$ 
Then we perform qq-plots of 
$$\eta_{D_{1}}^{(1)}-\mu_{D_{1}},\ldots,\eta_{D_{m}}^{(m)}-\mu_{D_{m}}$$
versus the standard Gumbel distribution.
As a measure of variability of the estimates, non-parametric block bootstrap methods (cf.~\citet{Politis5}, Section~3.2) are applied to obtain $95\%$ pointwise confidence bounds. Using bootstrap methods, we preserve the dependence between different subsets $D$ in the confidence intervals. Under $H_0$, the bisecting line should lie within these confidence bounds.

\begin{figure}[t]
\centering
\subfloat[]{\includegraphics[height=6cm,width=6cm]{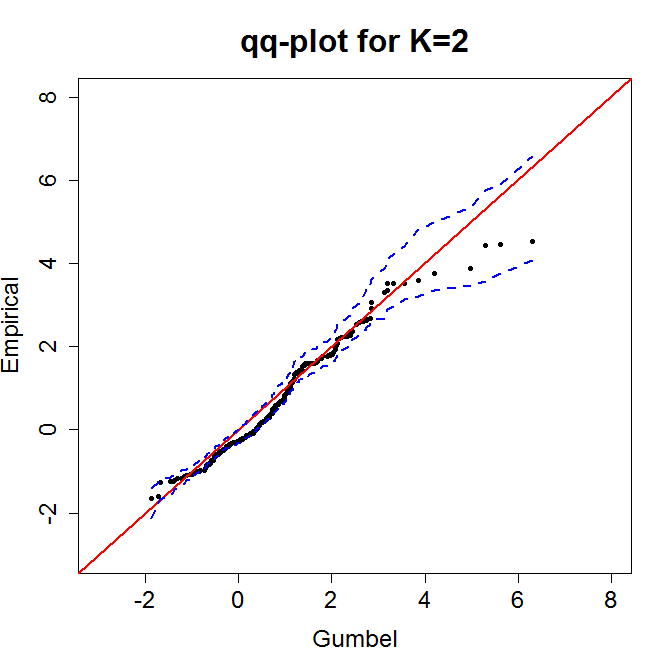}}
\subfloat[]{\includegraphics[height=6cm,width=6cm]{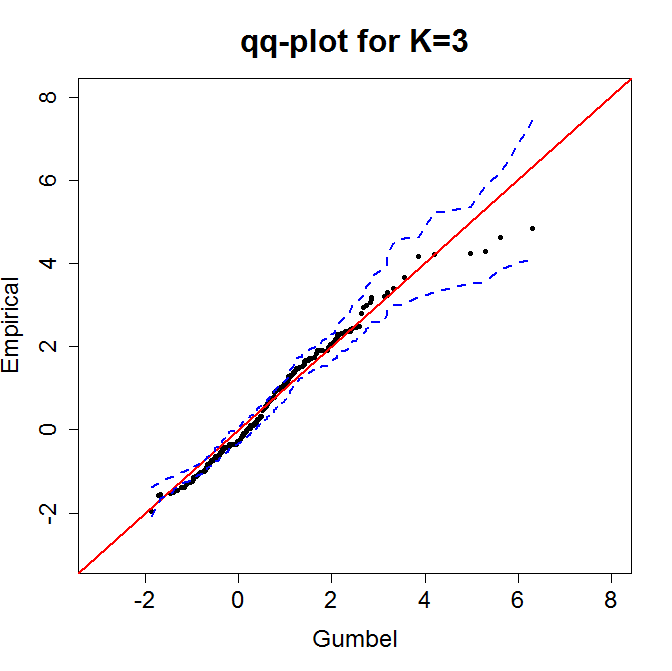}} \\
\subfloat[]{\includegraphics[height=6cm,width=6cm]{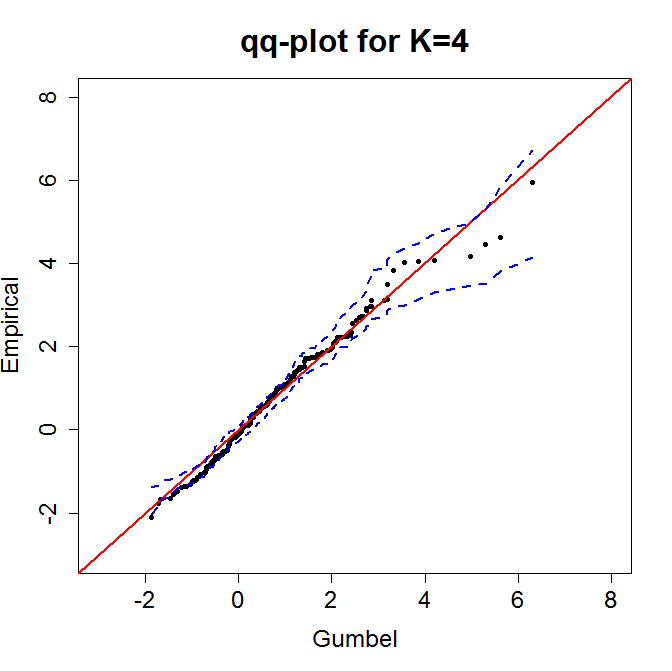}}
\subfloat[]{\includegraphics[height=6cm,width=6cm]{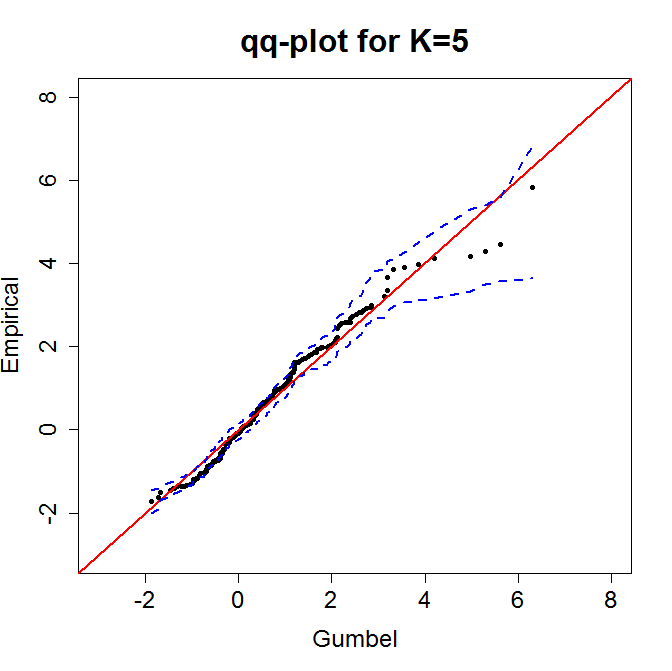}}
\caption{qq-plots of theoretical standard Gumbel quantiles versus the empirical quantiles (black dots). The latter correspond to the empirical distribution of maxima taken over groups of cardinality $K$. Dashed blue lines mark $95\%$ pointwise confidence bounds obtained by block bootstrap. Solid red lines correspond to no deviation.}
\label{Tawn_orig}
\end{figure}

The Florida daily rainfall maxima show only little temporal dependence beyond one day. Hence we choose $B_1=2$ and $B_2=1$, {which yields $R=\lfloor{\frac{732}{3}}\rfloor=244$ mutually independent time blocks of spatial data. We perform the described procedure for $K=2,3,4,5$, which entails $m=R=244$. Thus we obtain a total number of $N=244^2=59\,536$  subsets.}
The power of this diagnostic test increases with $K$ (cf. \citet{Wadsworth})  
as it gets less likely to include sets of space-time points that are $K$-wise independent. Figure~\ref{Tawn_orig} shows the results for the different choices of $K$. The solid red bisecting lines lie inside the confidence bounds. 
Hence, there is no statistically significant evidence of the space-time process generating the data not to be max-stable.

\subsection{Pairwise maximum likelihood estimation} \label{s53}

We apply the pairwise maximum likelihood estimation to the standard Fr\'echet transformed data \eqref{DataFrechet}. The parameters to estimate are those of the function $\delta$ in \eqref{3D}; i.e.,  $C_1, C_2, C_3\in (0,\infty)$ and $\alpha_1,\alpha_2,\alpha_3 \in (0,2]$.

{In the definition of the pairwise log-likelihood function \eqref{PWLgeneraldimension}, the maximum spatial and temporal lags are specified by the numbers $r_1$, $r_2$ and $p$, respectively. 
Immediately by model \eqref{3D} for $\delta$, the parameters of the three different dimensions (space and time) are separated in the extremal setting. 
This has also been noticed in \citet{Steinkohl2}, where a simulation study  in Section~7 for the isotropic model shows that estimating the {spatial and temporal} parameter pairs individually leads to very good results in terms of root-mean-square error and mean absolute error.
Hence, for example for parameter estimates for $C_1$ and $\alpha_1$, we can set the maximum lags corresponding to the remaining parameters equal to 0 (i.e., we set $r_2=p=0$).
This means that we basically fit univariate models to the respective spatial and temporal parts of the dependence function \eqref{3D}. Hence, this separation simplifies the statistical estimation. However, proving asymptotic properties of the pairwise likelihood estimator in the special case of a univariate model would for instance still involve showing the required mixing conditions and thus not remove much of the complexity.

Furthermore, we know that we should not include too many lags in space or time into the likelihood, since independence effects can introduce a bias in the estimates, see for example \citet{Nott}, Section~2.1, or \citet{Huser}, Section 4. 
On the other hand, an empirical analysis showed that extremal spatial dependence of the Florida daily rainfall maxima ranges up to lag 4 and extremal temporal dependence does not last more than one or two days, cf. Figure~7.2.6 in \citet{steinkohlphd}.
Hence, we perform the PMLE for maximum spatial and temporal lags up to 4 and 2, respectively, thus also assuring identifiability of all parameters according to Table~\ref{Table_Identifiability}. The results are summarised in Table~\ref{Florida_results}. 
Setting $r_1,$ $r_2$ or $p$ equal to 1 results in non-identifiability of the corresponding parameters $\alpha_1,$ $\alpha_2$ or $\alpha_3$, respectively; cf. Table~\ref{Table_Identifiability}. 
Therefore, they are not shown in Table~\ref{Florida_results}.

The combination of a rather large estimate for $\wh{C}_3$ and a rather small estimate for $\wh{\al}_3$ indicates that there is only little extremal temporal dependence, see \citet{steinkohlphd}, Section~7.2. Asymptotic $95\%$-confidence intervals are based on asymptotic normality of the parameter estimates and estimated using subsampling methods (cf. Section~\ref{s4}).

\begin{table}[t] 
\centering
\begin{tabular}{|p{2cm}|p{2.5cm}|p{2.5cm}|} 
\hline
\vspace*{0.001cm} \textbf{max. lags} &\vspace*{0.001cm} \textbf{$\wh C_i$} & \vspace*{0.001cm} \textbf{$\wh{\alpha}_i$} \\
\hline
(2,0,0) & 0.6287 $[0.5928,0.6646]$ & 0.9437 $[0.9065,0.9808]$\\
\hline 
(3,0,0) & 0.6358 $[0.5989,0.6728]$ & 0.8599 $[0.8189,0.9009]$\\
\hline
(4,0,0) & {0.6438 $[0.6051,0.6825]$} & 0.8107 $[0.7690,0.8525]$\\
\hline
(0,2,0) & 0.7271 $[0.6492,0.8050]$ & 0.9517 $[0.8715,1.0320]$\\
\hline
(0,3,0) & 0.7370 $[0.6586,0.8154]$ & 0.8521 $[0.7737,0.9305]$\\
\hline
(0,4,0) & {0.7476 $[0.6677,0.8275]$} & {0.7931 $[0.7039,0.8822]$}\\
\hline
(0,0,2) & 4.8378 $[4.4282,5.2474]$ & 0.1981 $[0.0177,0.3784]$\\
\hline
\end{tabular}
\caption{Estimates of the parameter pairs $(C_1,\alpha_1)$, $(C_2,\alpha_2)$ and $(C_3,\alpha_3)$ for different maximum spatial and temporal lags. Intervals below the point estimates are asymptotic $95\%$-confidence bounds based on subsampling.}
\label{Florida_results}
\end{table}

\subsection{Isotropic versus anisotropic model}\label{s54}

Using the results of Section~\ref{s4}, we want to apply the test \eqref{test} for spatial isotropy to the hypothesis
\begin{align}
H_0: \{C_1=C_2\,\mbox{ and } \, \al_1=\al_2\} \quad\mbox{versus}\quad
H_1: \{C_1\neq C_2\,\mbox{ or }\, \al_1\neq \al_2\}. \notag
\end{align}\noindent

For the block maxima of the precipitation data we have $d=2$, $M=12$ and $T=732$.
This corresponds to the situation of a fixed spatial domain with $\tau_{T}=\sqrt{732}$. 

We use the spatial PMLEs based on maximum lags 2-4, which can be read off from Table~\ref{Florida_results}.
We obtain the rejection areas from Theorem~\ref{subsampling_confFixed}.
We choose $b_1=b_2=5$, thus ensuring that the full range of spatial dependence is contained in the subsamples and simultaneously achieving that their number is large.
Concerning the number of time points in each subsample, we take $b_3=600$. Here we choose a large number to ensure that Theorem~\ref{subsampling_confFixed}, where $T \rightarrow \infty$, is applicable. This results in 
$\tau_{b_3}=\sqrt{b_3}=\sqrt{600}.$ In order to obtain a large number of subsamples, we further choose $e_1=e_2=e_3=1$ as the degree of overlap. 

\begin{table}[H] 
\centering
\begin{tabular}{|p{0.8cm}|p{1.2cm}|p{1.2cm}|p{2.3cm}|p{3.1cm}|p{2.4cm}|p{1.8cm}|} 
\hline
\textbf{max. lag} & $\tau_{T}$ & $\wh{C}_2-\wh{C}_1$  & $\tau_{T}(\wh{C}_2-\wh{C}_1)$ & Rej$_{\wh \theta_C}^{(T)}$  & \textbf{97.5\%-CI for} $C_2-C_1$ & \textbf{Reject $C_1=C_2$}\\
\hline
2 & 27.055 & 0.098 & 2.651 & $[-2.400, 2.400]^c$ & $[0.010,0.187]$ & yes\\
\hline
3 & 27.055 & 0.101 & 2.738 & $[-2.392, 2.392]^c$ & $[0.013,0.190]$ & yes\\
\hline
{4} & 27.055 & 0.104 & 2.808 & $[-2.393,2.393]^c$ & $[0.015,0.192]$ & yes\\
\hline
\end{tabular}
\caption{Test results for parameters $C_1$ and $C_2$. All values are rounded to three positions after decimal point.}
\label{Test_result_C}
\end{table}
\begin{table}[h] 
\centering
\begin{tabular}{|p{0.8cm}|p{1.2cm}|p{1.2cm}|p{2.3cm}|p{3.1cm}|p{2.4cm}|p{1.8cm}|} 
\hline
\textbf{max. lag} & $\tau_{T}$ & \textbf{$\wh{\alpha}_2-\wh{\alpha}_1$} & $\tau_{T}(\wh{\alpha}_2-\wh{\alpha}_1)$ & $\text{Rej}_{\wh \theta_{\alpha}}^{(T)}$  & \textbf{97.5\%-CI for} $(\alpha_2-\alpha_1)$ & \textbf{Reject $\alpha_1=\alpha_2$}\\
\hline
2 & 27.055 & 0.008 & 0.216 & $[-2.162, 2.162]^c $ & $[-0.072,0.088]$ & no\\
\hline
3 & 27.055 & -0.008 & -0.216 & $[-2.130, 2.130]^c $ & $[-0.087,0.071]$ & no\\
\hline
4 & 27.055 & -0.018 & -0.477 & $[-2.342, 2.342]^c$ & $[-0.104,0.069]$ & no\\
\hline
\end{tabular}
\caption{Test results for parameters $\alpha_1$ and $\alpha_2$. All values are rounded to three positions after decimal point.}
\label{Test_result_alpha}
\end{table}

Tables~\ref{Test_result_C} and~\ref{Test_result_alpha} present the results of the two tests at individual confidence levels ${\beta}=2.5\%$ giving a test for \eqref{test} at a confidence level $2\beta=5\%$ by Bonferroni's inequality. The differences $(\wh{C}_2-\wh{C}_1)$ and $(\wh{\alpha}_2-\wh{\alpha}_1)$ can be obtained from Table~\ref{Florida_results}. 

Since we can reject the individual hypothesis that $C_1=C_2$ at a confidence level of 2.5$\%$, we can reject the overall hypothesis $H_0$ of \eqref{test} at a confidence level of $5\%$ and conclude that our data originate from a spatially anisotropic max-stable Brown-Resnick process. 
Further note the interesting fact that, although the asymptotic confidence interval for the difference $C_2-C_1$ does not include 0, the individual intervals for $C_1$ and $C_2$ overlap, see Table~\ref{Florida_results}. 
This is due to the fact that the individual confidence bounds are estimated independently of each other, whereas the estimated bounds for the difference reflect how far the parameter estimates lie apart in one fixed particular (sub)sample.

\subsection{Model check}\label{s55}

Finally, having fitted the Brown-Resnick space-time model \eqref{BR} to the precipitation data, we want to assess the quality of the fit. We take inspiration from Section~\ref{s52} of \citet{DavisonPadoanRibatet} and compare maxima taken over subsets of the space-time precipitation data with simulated counterparts.

Similarly as in Section~\ref{s52}, we consider subsets of the observations on a regular grid for $L$ spatial locations and for time points $1,\ldots, B_1$,
$$D=\{(s_1^{(\ell)},s_2^{(\ell)},1),\ldots,(s_1^{(\ell)},s_2^{(\ell)},B_1): \ell=1,\ldots,L\}.$$
We follow the procedure as in \eqref{blocks} to extract $R$ independent realisations of 
$\{\eta_1(\bs{s},t): (\bs{s},t) \in D\}$ from 
the standard Gumbel transformed space-time observations \eqref{DataGumbel2}. 
This yields in turn $R$ independent realisations of 
$\eta_D=\max \{\eta_1(\bs{s},t): (\bs{s},t) \in D\}$, which we summarise in the {ordered vector} $\eta_{\text{data}}:=(\eta_D^{(1)},\ldots,\eta_D^{(R)}).$ Now we simulate a corresponding {vector}, denoted by  $\wh\eta_{\text{sim}}:=(\wh\eta_D^{(1)},\ldots,\wh\eta_D^{(R)}).$
To this end we need reliable Monte Carlo values as elements of $\wh\eta_{\text{sim}}$.
We obtain them by simulating empirical order statistics as follows.
We simulate $m \cdot R$ independent copies of the Brown-Resnick space-time process on $D$ with dependence structure $\delta$ as in \eqref{vario0} with the PMLEs from Table~\ref{Florida_results}, 
where we take the estimates based on maximum lag {4} (for the spatial parameters) and 2 (for the temporal parameters), which are the maximum lags, where dependence is still present.
We transform the univariate margins to standard Gumbel.
This results in corresponding $m\cdot R$ independent simulations of $\eta_D$ and we consider them as $m$ blocks of size $R$. 
We order the $R$ values in each block and define $\wh{\eta}_D^{(i)}$ as the mean of all simulated $i$th order statistics for $i  =1,\ldots, R$, which gives $\wh{\eta}_{\text{sim}}:=(\wh{\eta}_D^{(1)},\ldots,\wh{\eta}_D^{(R)}).$

\begin{figure}[h]
\centering
\subfloat[]{\includegraphics[height=6cm,width=6cm]{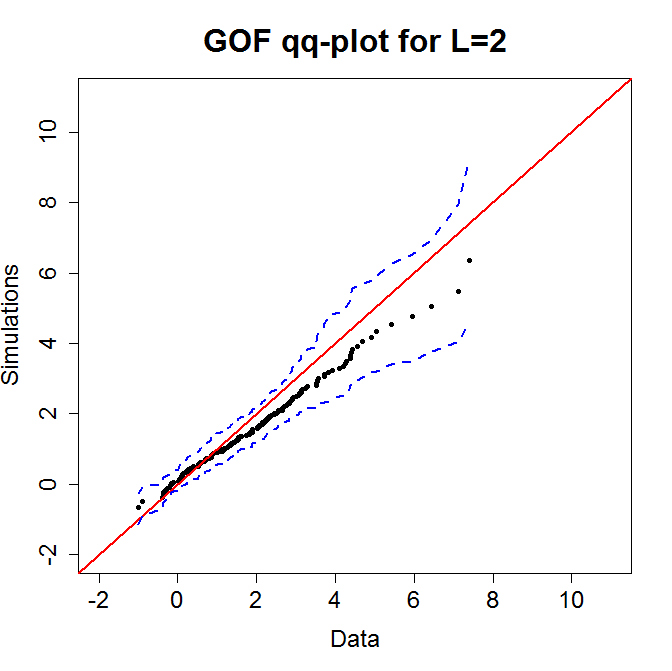}}
\subfloat[]{\includegraphics[height=6cm,width=6cm]{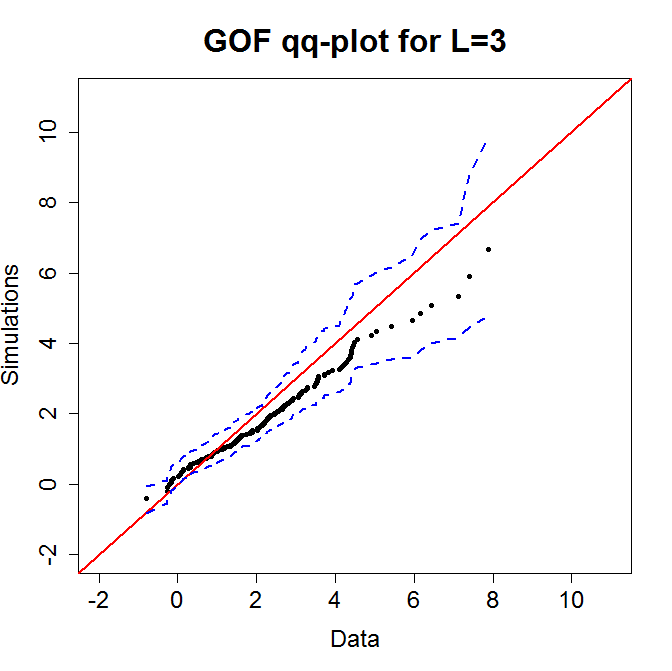}} \\
\subfloat[]{\includegraphics[height=6cm,width=6cm]{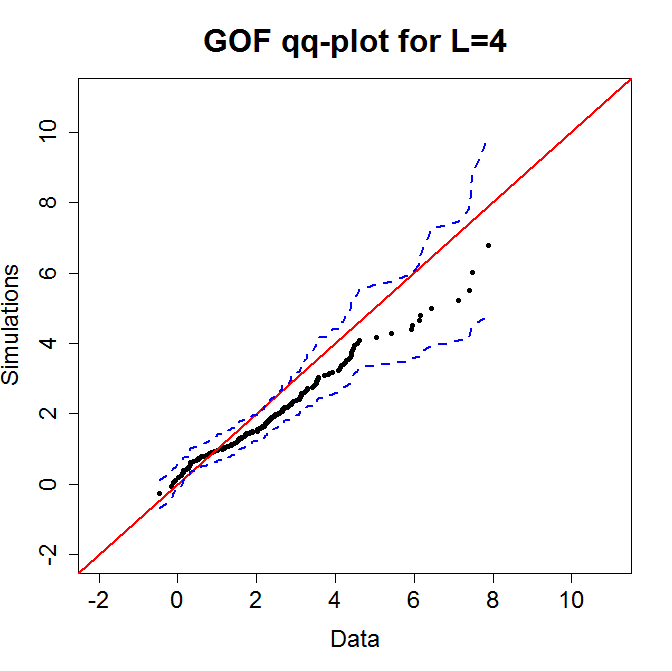}}
\subfloat[]{\includegraphics[height=6cm,width=6cm]{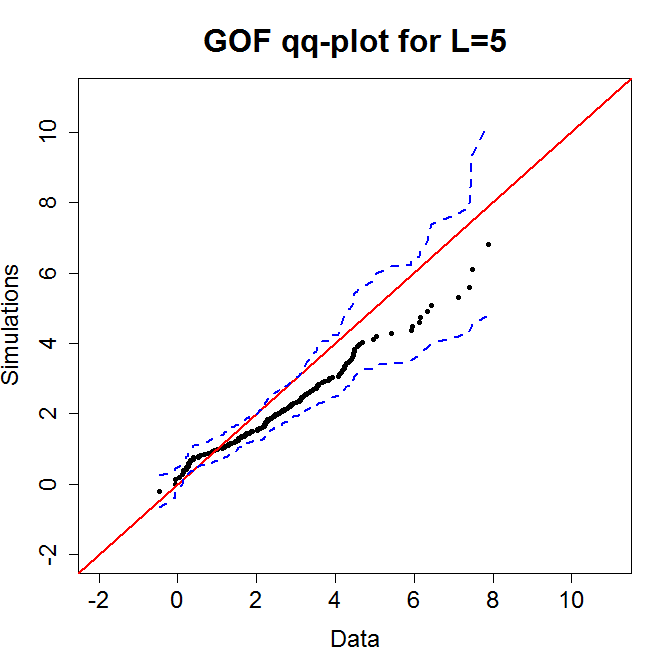}}
\caption{Goodness of fit qq-plots for different spatial locations and different $L$. Top left: $L=2$: (1,1) and (1,2). Top right: $L=3$: (1,1), (1,2) and (3,1). Bottom left: $L=4$: (1,1), (1,2), (3,1) and (3,2). Bottom right: $L=5$: (1,1), (1,2), (3,1), (3,2) and (2,1). 
PMLEs underlying the simulations are based on maximum spatial and temporal lags 4 and 2, respectively. Dashed blue lines mark 95$\%$ pointwise confidence bounds. Solid red lines correspond to no deviation.}
\label{gof}
\end{figure}

The {vectors} $\eta_{\text{data}}$ and $\wh{\eta}_{\text{sim}}$ are compared by qq-plots. If the fit is good, the points in the plots lie approximately on the bisecting line. Pointwise $95\%$-confidence bands are determined by the $2.5\%$ and the $97.5\%$ quantiles of the simulated order statistics. As in Section~\ref{s53}, we choose $B_1=2$. The number of simulations is $N=m \cdot R=100 \cdot 244=24400.$ Figure \ref{gof} presents the results for four exemplary groups of locations. The plots reveal a good model fit. 

We carried out the simulations using the exact method recently suggested in \citet{Dombry2}, Sections~3.3 and 5.2. For an overview and comparison of different simulation methods for Brown-Resnick processes we refer to \citet{Leber}.

\subsection{Application: conditional probability fields}

\begin{figure}[t] \centering
\subfloat[]{\includegraphics[height=4.6cm,width=6cm]{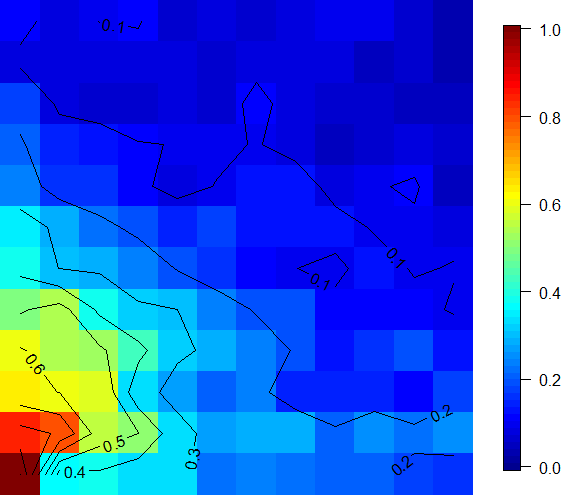}} \hspace*{0.3cm}
\subfloat[]{\includegraphics[height=4.6cm,width=6cm]{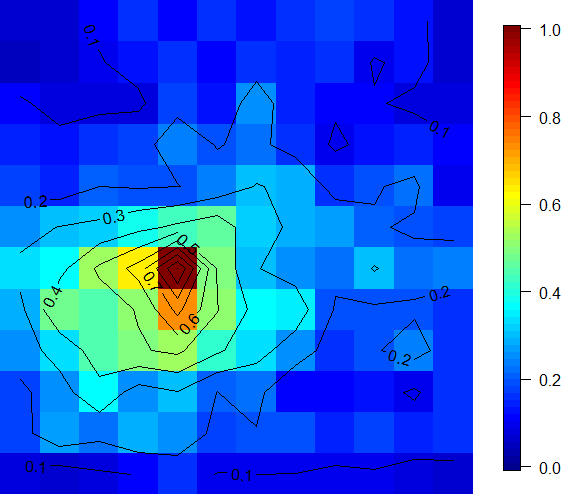}} \\
\subfloat[]{\includegraphics[height=4.6cm,width=6cm]{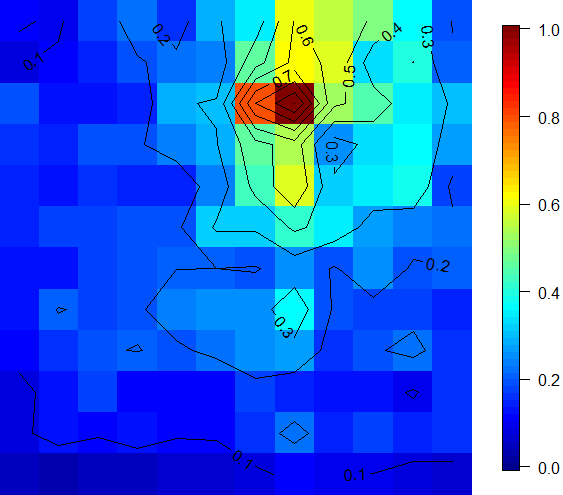}} \hspace*{0.3cm}
\subfloat[]{\includegraphics[height=4.6cm,width=6cm]{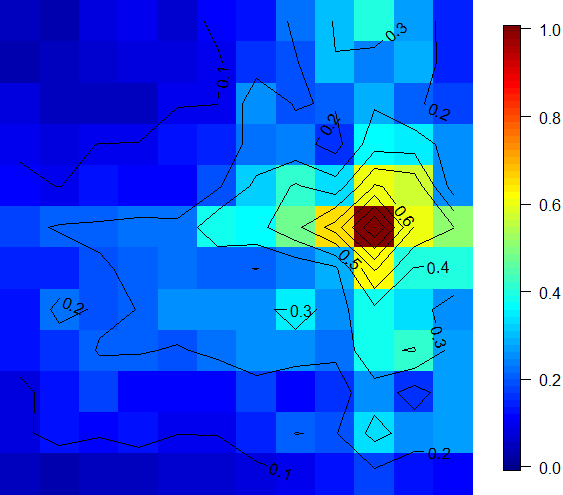}}
\caption{Predicted conditional probability fields based on daily maxima for reference space-time points (1,1,1), (5,6,1), (8,10,1) and (10,7,1) and rainfall levels $z=z^{\star}=2.5$ (clockwise from the top left to the bottom right).}
\label{cond_prob_field}
\end{figure}

Based on the fitted model, we want to answer questions like: Given there is extreme rain at some space-time reference point $(s_1^{\star},s_2^{\star},t^{\star}) \in \{1,\ldots,12\}^2 \times \{1,\ldots,732\}$, what is the estimated probability of extreme rain at some prediction space-time point $(s_1^p,s_2^p,t^p)$? In other words, we want to estimate the probabilities
\begin{align}
\mathbb{P}\left(\tilde{\eta}((s_1^p,s_2^p),t^p)>z\mid \tilde{\eta}((s_1^{\star},s_2^{\star}),t^{\star})>z^{\star}\right), \label{condprob}
\end{align}
where $\{\tilde{\eta}((s_1,s_2),t): s_1,s_2=1,\ldots,12, t=1, \ldots, 732\}$ are the stationary observations \eqref{DataGumbel1} and $z$ and $z^{\star}$ are prediction and reference rainfall levels, respectively. Denote by $\Lambda_{\mu,\sigma}$ the Gumbel distribution with location and scale parameters $\mu$ and $\sigma$ (cf. Section \ref{s51}) and set $\widehat{\mu}^p:=\widehat{\mu}(s_1^p,s_2^p)$, $\widehat{\sigma}^p:=\widehat{\sigma}(s_1^p,s_2^p)$, $\widehat{\mu}^{\star}:=\widehat{\mu}(s_1^{\star},s_2^{\star})$ and $\widehat{\sigma}^{\star}:=\widehat{\sigma}(s_1^{\star},s_2^{\star})$, which are the marginal Gumbel parameter estimates. Simple computations show that \eqref{condprob} can be estimated by 
\begin{align*}
&\frac{1}{1-\Lambda_{\widehat{\mu}^{\star},\widehat{\sigma}^{\star}}(z^{\star})}
\Big(1-\Lambda_{\widehat{\mu}^{\star},\widehat{\sigma}^{\star}}(z^{\star})- \Lambda_{\widehat{\mu}^p,\widehat{\sigma}^p}(z)\\
&+\exp\Big\{-\widehat{V}_D\Big(-\frac{1}{\log\left\{\Lambda_{\widehat{\mu}^p,\widehat{\sigma}^p}(z)\right\}},-\frac{1}{\log\left\{\Lambda_{\widehat{\mu}^{\star},\widehat{\sigma}^{\star}}(z^{\star})\right\}}\Big)\Big\} \Big),
\end{align*}

where $\widehat{V}_D$ is the estimate of the exponent measure \eqref{exponentmeasure} obtained by plugging in the PMLEs of the parameters of the dependence function $\delta$. Figure \ref{cond_prob_field} shows four predicted conditional probability fields for the reference points $(1,1,1)$, $(5,6,1)$, $(8,10,1)$ and $(10,7,1)$ and for high empirical rainfall levels $z=z^{\star}=2.5$. Because of the little temporal dependence in the daily maxima, we only consider equal time points for spatial predictions.

\subsection*{Acknowledgements}
We take pleasure in thanking Anthony Davison and his group for an extremely pleasant and interesting time at EPFL Lausanne, and SB also at a Summer School in Leukerbad.
SB also thanks Christina Steinkohl for her constant support, when writing his Master's Thesis. Discussions with Richard Davis {and Jenny Wadsworth} are gratefully acknowledged.
We also thank Chin Man Mok for providing the Florida rainfall data and acknowledge that the data are provided by the Southwest Florida Water Management District (SWFWMD). 
We thank the referees for fruitful comments and remarks.
SB additionally acknowledges that he was supported by Deutsche Forschungsgemeinschaft (DFG) through the TUM  International Graduate School of Science and Engineering (IGSSE).

\bibliography{bibtex_spacetime}
\bibliographystyle{abbrvnat}

\appendix
\section{An auxiliary lemma}\label{a1}

\ble\label{aux}
The following two bounds hold true for $r \geq 1$, $\alpha \in (0,2]$ and $C>0$:
\beam
\label{formela1} & \int\limits_y^{\infty}u^r\ex{-C u^{\alpha}} \,\mathrm{d}u \sim  \frac{1}{C\alpha} y^{r-\alpha+1} \ex{-Cy^{\alpha}},\quad y\to\infty,\\
\label{mixint} & \int\limits_1^{\infty}\Big(\int\limits_y^{\infty} u^r \ex{-Cu^{\alpha}} \,\mathrm{d}u \Big)^{\frac{1}{3}} \,\mathrm{d}y <\infty.
\eeam
\ele

\bproof
First note that integrals of the form $\int_0^{\infty} u^{r} \ex{-C u^{\alpha}} \,\mathrm{d}u$ are finite for every $r > -1$, $\alpha \in (0,2]$, and $C>0$, since they are transformations of the gamma function $\Gamma(x)=\int_0^{\infty} t^{x-1}\ex{-t}\,\mathrm{d}t$, which exists for positive $x$. 
We prove \eqref{formela1} by an application of l'H\^{o}pital's rule:
\begin{align*}
\lim\limits_{y \rightarrow \infty} \frac{\int_y^{\infty}u^r\ex{-Cu^{\alpha}} \,\mathrm{d}u}{\frac{1}{C\alpha} y^{r-\alpha+1} \ex{-Cy^{\alpha}}}=\lim\limits_{y \rightarrow \infty}\frac{-y^r\ex{-Cy^{\alpha}}}{\left(-y^r+\frac{r-\alpha+1}{C\alpha}y^{r-\alpha}\right)\ex{-Cy^{\alpha}}}=\lim\limits_{y \rightarrow \infty}\frac{y^r}{y^r\left(1-\frac{r-\alpha+1}{C\alpha}y^{-\alpha}\right)}=1.
\end{align*}
In order to prove \eqref{mixint}  first note that it follows from \eqref{formela1}  that for every $\epsilon>0$ there exists $y_0=y_0(\epsilon)$ such that for all $y \geq y_0$, 
\begin{align}\label{mixitasy}
\Big(\int\limits_y^{\infty}u^r\ex{-Cu^{\alpha}} \,\mathrm{d}u\Big)^{\frac{1}{3}} 
\leq (1+\epsilon)\Big(\frac{1}{C\alpha}\Big)^{\frac{1}{3}} y^{\frac{r-\alpha+1}{3}} \ex{-\frac{C}{3}y^{\alpha}}. 
\end{align}
Now we split the double integral of \eqref{mixint} up into
\begin{align*}
\int\limits_1^{y_0}\Big(\int\limits_y^{\infty} u^r \ex{-Cu^{\alpha}} \,\mathrm{d}u \Big)^{\frac{1}{3}} \,\mathrm{d}y + \int\limits_{y_0}^{\infty}\Big(\int\limits_y^{\infty} u^r \ex{-Cu^{\alpha}} \,\mathrm{d}u \Big)^{\frac{1}{3}} \,\mathrm{d}y=:I_1+I_2.
\end{align*}
For $I_1$ we obtain
$$I_1 \leq \int\limits_1^{y_0}\Big( \int\limits_y^{y_0} u^r \ex{-Cu^{\alpha}} \,\mathrm{d}u \Big)^{\frac{1}{3}} \,\mathrm{d}y + \int\limits_1^{y_0}\Big(\int\limits_{y_0}^{\infty} u^r \ex{-Cu^{\alpha}} \,\mathrm{d}u \Big)^{\frac{1}{3}} \,\mathrm{d}y
=:I_1^{(1)}+I_1^{(2)}.$$
$I_1^{(1)}$ is obviously finite, and to bound $I_1^{(2)}$ we use \eqref{mixitasy}, which yields
$$I_1^{(2)} \leq (y_0-1)(1+\epsilon)\Big(\frac{1}{C\alpha}\Big)^{\frac{1}{3}}y_0^{\frac{r-\alpha+1}{3}}\ex{-\frac{C}{3}y_0^{\alpha}} < \infty.$$
Concerning $I_2$, note that 
$$I_2 \leq (1+\epsilon) \Big(\frac{1}{C\alpha}\Big)^{\frac{1}{3}} \int\limits_{y_0}^{\infty} y^{\frac{r-\alpha+1}{3}}\ex{-\frac{C}{3}y^{\alpha}} \,\mathrm{d}y,$$
which is finite by finiteness of the gamma function.
\eproof

\end{document}